\documentclass[11pt,a4paper]{article}
\pdfoutput=1
\usepackage{latexsym}
\usepackage{chapterbib}
\usepackage{graphics,graphicx}
\usepackage{amssymb,amsmath,amsfonts,amstext}
\usepackage{hyperref}
\usepackage[bbgreekl]{mathbbol}
\usepackage{color,xcolor}
\usepackage{hyperref}
\usepackage{geometry}  
\usepackage{braket}
\usepackage{cancel}
\usepackage{float}
\usepackage[utf8]{inputenc}
\usepackage{setspace}
\def\){\right)}
\def\({\left( }
\def\]{\right] }
\def\[{\left[ }

\def\NO{\nonumber}

\newcommand{\be}{\begin{equation}}
\newcommand{\ee}{\end{equation}}

\def\bea{\begin{eqnarray}}
\def\eea{\end{eqnarray}}

\def\bal#1\eal{\begin{align}#1\end{align}}

\def\bald{\begin{aligned}}
\def\eald{\end{aligned}}

\def\bsub{\begin{subequations}}
\def\esub{\end{subequations}}

\def\beqx{\begin{displaymath}}
\def\eeqx{\end{displaymath}}

\newcommand{\bmat}{\left(\begin{array}}
\newcommand{\emat}{\end{array}\right)}

\def\half{\frac{1}{2}}

\def\a{\alpha}

\def\c{\chi}
\def\d{\delta}
\def\e{\epsilon}
\def\f{\phi}
\def\g{\gamma}

\def\k{\kappa}

\def\m{\mu}
\def\n{\nu}
\def\o{\omega}
    
\def\p{\pi}

    \def\th{\theta}
\def\r{\rho}
\def\s{\sigma}
\def\t{\tau}
\def\x{\xi}

\def\J{\Psi}

\def\O{\Omega}
    \def\Om{\Omega}
\def\P{\Pi}

    \def\Th{\Theta}
\def\S{\Sigma}

\def\ve{\varepsilon}

\def\vf{\varphi}

\def\ba{\bbalpha}
\def\bk{\bbkappa}

\def\bs{\bbsigma}

\def\ce{{\cal E}}
\def\cf{{\cal F}}
\def\cg{{\cal G}}
\def\ch{{\cal H}}

\def\cj{{\cal J}}
\def\ck{{\cal K}}

\def\cm{{\cal M}}
\def\cn{{\cal N}}
\def\co{{\cal O}}

\def\cq{{\cal Q}}

\def\cs{{\cal S}}
\def\ct{{\cal T}}

\def\cx{{\cal X}}
\def\cy{{\cal Y}}

\def\bb#1{\ensuremath{\mathbb{#1}}} 

\def\bo{{\raise-.3ex\hbox{\large$\Box$}}}               
\def\pa{\partial}                                       
\def\face{{\raise.2ex\hbox{$\displaystyle \bigodot$}\mskip-2.2mu \llap {$\ddot
        \smile$}}}                                   
\def\>{\rangle}                                      
\def\<{\langle}                                      

\def\tx#1{\text{#1}}
\def\sbtx#1{{}_{\rm #1}}                           
\def\wt#1{\widetilde{#1}}                            
\def\lbar#1{\ensuremath{\overline{#1}}}              
\def\leftrightarrowfill{$\mathsurround=0pt \mathord\leftarrow \mkern-6mu
        \cleaders\hbox{$\mkern-2mu \mathord- \mkern-2mu$}\hfill
        \mkern-6mu \mathord\rightarrow$}        
\def\dvec#1{\vbox{\ialign{##\crcr
        \leftrightarrowfill\crcr\noalign{\kern-1pt\nointerlineskip}
        $\hfil\displaystyle{#1}\hfil$\crcr}}}           

\def\-{\hphantom{-}}

\begin{document}

\begin{titlepage}

\pagestyle{empty}

\begin{flushright}
 {\small KIAS-P18004\\ UPR-1288-T}
\end{flushright}
\vskip1.5in

\begin{center}
\textbf{\Large Thermoelectric DC conductivities in hyperscaling violating Lifshitz theories \rule{0cm}{.5cm}}
\end{center}
\vskip0.2in

\begin{center}
{\large 
Sera Cremonini$^{a,}$\footnote{\href{mailto: cremonini@lehigh.edu}{\tt cremonini@lehigh.edu}}\thinspace , 
Mirjam Cveti\v c$^{b,c,}$\footnote{\href{mailto: cvetic@physics.upenn.edu}{\tt cvetic@physics.upenn.edu}}
\thinspace ,  
Ioannis Papadimitriou$^{d,}$\footnote{\href{mailto: ioannis@kias.re.kr}{\tt ioannis@kias.re.kr}}}
\end{center}
\vskip0.2in

\begin{center}
{\small {$^{a}$}Department of Physics, Lehigh University,
	Bethlehem, PA 18018, USA}\\ \vskip0.1in
{\small {$^{b}$}Department of Physics and Astronomy, University of Pennsylvania,\\ Philadelphia, PA 19104-6396, USA}\\ \vskip0.1in
{\small {$^{c}$}Center for Applied Mathematics and Theoretical Physics,\\
University of Maribor, SI2000 Maribor, Slovenia}\\ \vskip0.1in
{\small {$^{d}$}School of Physics, Korea Institute for Advanced Study, Seoul 02455, Korea}
\end{center}
\vskip0.2in

\begin{abstract}
We analytically compute the thermoelectric conductivities at zero frequency (DC) in the holographic dual of a four dimensional Einstein-Maxwell-Axion-Dilaton theory that admits a class of asymptotically hyperscaling violating Lifshitz backgrounds with a dynamical exponent $z$ and hyperscaling violating parameter $\th$. We show that the heat current in the dual Lifshitz theory involves the energy flux, which is an irrelevant operator for $z>1$. The linearized fluctuations relevant for computing the thermoelectric conductivities turn on a source for this irrelevant operator, leading to several novel and non-trivial aspects in the holographic renormalization procedure and the identification of the physical observables in the dual theory. Moreover, imposing Dirichlet or Neumann boundary conditions on the spatial components of one of the two Maxwell fields present leads to different thermoelectric conductivities. Dirichlet boundary conditions reproduce the thermoelectric DC conductivities obtained from the near horizon analysis of Donos and Gauntlett, while Neumann boundary conditions result in a new set of DC conductivities. We make preliminary analytical estimates for the temperature behavior of the thermoelectric matrix in appropriate regions of parameter space. In particular, at large temperatures we find that the only case which could lead to a linear resistivity $\rho \sim T$ corresponds to $z=4/3$.   
\end{abstract}

\end{titlepage}

\tableofcontents
\addtocontents{toc}{\protect\setcounter{tocdepth}{3}}
\renewcommand{\theequation}{\arabic{section}.\arabic{equation}}
	
\newpage
\section{Introduction and summary of results}
\setcounter{equation}{0}

Recent years have seen growing efforts to extend the dictionary of holography beyond the original paradigm of the AdS/CFT correspondence, and generalize it to spacetimes with different asymptotics from those of AdS. In particular, there has been interest in extensions to quantum field theories that may not be relativistic in the UV, with Lifshitz theories providing a prime example \cite{Kachru:2008yh,Taylor:2008tg,Ross:2009ar,Ross:2011gu,Mann:2011hg,Baggio:2011cp,Griffin:2011xs,Griffin:2012qx,Chemissany:2014xsa}. These efforts are part of a broader program to enlarge the universality class of theories to which holographic techniques can be applied, and also to understand how symmetries and symmetry breaking mechanisms are encoded in a spacetime description. Much of this program has been motivated by the intriguing behavior of many novel, strongly correlated quantum phases of matter that typically lack a quasiparticle description, but are amenable to being probed with the tools of holography. Indeed, the latter can offer a new analytical window into the dynamics of such phases, and provide insight into their unconventional transport behavior. For a comprehensive review with a focus on condensed matter applications see \emph{e.g.}, \cite{Hartnoll:2016apf}.

Thus far, most holographic studies of transport have been framed within the context of systems
which break Lorentz invariance in the IR, but emanate from a relativistic conformal fixed point in the UV. Such constructions, which rely on geometries that approach AdS at the boundary, have proven fruitful for probing a rich variety of IR phases, whose behavior is captured by the near-horizon region of the geometry. However, non-relativistic UV fixed points and their hyperscaling violating generalizations are interesting in their own right, and furthermore broaden the class of renormalization group (RG) flows that can be modeled holographically, offering novel ways to make contact with condensed matter systems. In particular, in order to identify the physical observables in non-relativistic theories holographically it is necessary to place them at the UV of the RG flow, that is to consider bulk backgrounds that are asymptotically non-relativistic, and develop the corresponding holographic dictionary.   

To this end, in this paper we examine a four dimensional Einstein-Maxwell-Axion-Dilaton theory that admits a class of non-relativistic geometries that are asymptotically Lifshitz and exhibit hyperscaling violation. They are parametrized by a dynamical critical exponent $z$ and a hyperscaling violating parameter $\theta$, which characterizes the anomalous scaling of the free energy of the system. The axionic fields included in the model break translational invariance along the boundary directions, thus providing a mechanism to dissipate momentum \cite{Karch:2007pd,Hartnoll:2009ns,Faulkner:2010zz,Faulkner:2013bna}, a crucial ingredient for a realistic description of materials with impurities and an underlying lattice structure. The theory we focus on involves two massless $U(1)$ gauge fields, one responsible for the Lifshitz-like nature of the background solutions, while the other is analogous to a standard Maxwell field in asymptotically-AdS charged black holes.

The electrical DC conductivity matrix for this model was computed originally
in \cite{Cremonini:2016avj}, using the near horizon analysis of \cite{Donos:2014cya}. Subsequently, \cite{Bhatnagar:2017twr} computed the full thermoelectric conductivity matrix in the presence of a background magnetic field, using the same near horizon analysis.\footnote{See also \cite{Ge:2016lyn,Chen:2017gsl}. Note, however, that there are a number of subtleties in their analysis, including choices of parameters which lead to violations of the null energy condition. Some of their results differ from ours.} 
However, the near horizon analysis does not suffice to identify the conserved currents in the dual Lifshitz theory whose two-point functions determine the thermoelectric conductivities. Moreover, both the identification of the conserved currents in the Lifshitz theory and the corresponding conductivities depend on the boundary conditions for the bulk fields imposed at infinity. In particular, the conductivities obtained from the near horizon analysis correspond only to one specific boundary condition, out of an infinite family of possible boundary conditions.   

In this paper we consider linearized fluctuations around a class of purely electric asymptotically hyperscaling violating Lifshitz backgrounds with axion charge that turn on all the modes necessary to obtain the full matrix of thermoelectric conductivities. Through a careful analysis of the asymptotic behavior of the solutions at the UV we determine the boundary counterterms necessary to renormalize the theory and identify the physical observables in the dual Lifshitz theory. One of our main observations is that the fluctuations necessary to compute the thermoelectric conductivities turn on a source for the energy flux, which is an irrelevant operator in the energy-momentum complex of the dual Lifshitz theory \cite{Ross:2011gu}. This leads to several subtleties in the holographic renormalization of the theory, which we address in detail. Moreover, we consider both Dirichlet and Neumann boundary conditions for the gauge field supporting the Lifshitz background and show that these boundary conditions result in a different set of thermoelectric conductivities for the second Maxwell field. In fact, there is an infinite set of possible boundary conditions, parameterized by $SL(2,\bb Z)$ \cite{Witten:2003ya}, all leading to a different set of conductivities. Only the conductivities corresponding to Dirichlet boundary conditions on the gauge field supporting the Lifshitz background match those obtained from the near horizon analysis, which has mostly been used in the literature. However, we show that boundary conditions at the UV provide a mechanism for obtaining different conductivities from the same bulk theory.     

The conductivities we obtain from Dirichlet and Neumann boundary conditions have a rich but distinct behavior as a function of temperature and are sensitive to the various parameters characterizing the background solution. We perform a preliminary analysis of their temperature dependence by looking for regions in the parameter space where they exhibit approximate scaling behavior with the temperature. We identify several clean scaling regimes in the limit of large temperature, which in our setup probes the Lifshitz theory in the UV. Rather intriguingly, the only case we could identify that can potentially lead to a linear resistivity arises in the case of Dirichlet boundary conditions for the specific value $z=4/3$ of the dynamical exponent, which was also singled out in the field theory analysis of \cite{Hartnoll:2015sea}.

The paper is organized as follows. In section \ref{sec:model} we present the general class of models we are interested in, both in the Einstein and the so called `dual frame' \cite{Kanitscheider:2009as}, as well as the generic backgrounds around which we linearize the equations of motion. We consider backgrounds carrying only electric Maxwell charge and magnetic axion charge in order to incorporate momentum dissipation. A significant part of our analysis is carried out for generic backgrounds and applies to a wide range of UV asymptotics, including AdS$_4$ and hyperscaling violating Lifshitz with arbitrary dynamical exponents. A specific exact asymptotically hyperscaling violating Lifshitz black brane solution studied in \cite{Cremonini:2016avj} is reviewed in subsection \ref{example} and is the focus of most of our subsequent analysis. Section \ref{sec:linear} contains our study of the linearized fluctuation equations, which are collected in appendix \ref{sec:fluctuations}. In particular, we obtain the general solution of the fluctuation equations in the small frequency limit and with infalling boundary conditions on the horizon. Specializing to the exact background of subsection \ref{example}, we then obtain the asymptotic UV expansions of the linear fluctuations and determine the boundary terms required in order to renormalize the theory. Finally, we identify the physical observables in the dual Lifshitz theory for two different boundary conditions on the bulk gauge field supporting the Lifshitz background. All two-point functions captured by the linear fluctuations we consider are presented in section \ref{sec:conductivities}, and the corresponding DC conductivities are obtained after the identification of the thermal current in the dual Lifshitz theory. We conclude in section \ref{sec:conclusion}, where we also discuss a number of possible directions for future work. The radial Hamiltonian description of the bulk dynamics, which is used to renormalize the theory, is summarized in appendix \ref{ham}.

\section{Hyperscaling violating Lifshitz black branes with axion charge}
\label{sec:model}
\setcounter{equation}{0}

We are interested in asymptotically hyperscaling violating Lifshitz backgrounds supported by a massless gauge field and a running dilaton \cite{Taylor:2008tg}. Additionally, we turn on spatial profiles for a number of axion fields in order to break translation invariance \cite{Donos:2013eha,Andrade:2013gsa}, which ensures that the DC conductivities are finite.

\paragraph{Einstein frame} Specifically, the model we consider is described by the Einstein frame action
\be\label{action-0}
S=\frac{1}{2\k^2}\int_\cm d^{d_s+2}x\sqrt{-g}\Big(R-\a (\pa\f)^2-\S_{IJ}(\f)F^I_{\m\n}F^{J\m\n}-Z(\f)(\pa\c^a)^2-V(\f)\Big)+S\sbtx{GH},
\ee
where $d_s$ denotes the number of spatial dimensions of the conformal boundary, $\k^2=8\p G_{d_s+2}$ is the gravitational constant in $d_s+2$ dimensions, and $S\sbtx{GH}$ denotes the Gibbons-Hawking term 
\be
S\sbtx{GH}=\frac{1}{2\k^2}\int_{\pa\cm} d^{d_s+1}x\sqrt{-\g}\;2K.
\ee
We have allowed for a generic normalization of the dilaton kinetic term, corresponding to the positive definite parameter $\a$, for later convenience. The indices $a=1,\cdots, d_s$ run over all $d_s$ axion fields, while the indices $I,J$ run over a yet unspecified number of Abelian gauge fields. The summation convention is adopted for the indices $a$ and $I, J$, in addition to the spacetime indices $\m, \n$. Finally, the symmetric matrix $\S_{IJ}(\f)$ and the functions $Z(\f)$ and $V(\f)$ are going to be kept generic for most part of our analysis, except that $\S_{IJ}(\f)$ is required to be invertible with strictly positive eigenvalues. 

The equations of motion following from the action \eqref{action-0} are
\bal\label{eoms-0}
&R_{\m\n}=Z(\f)\pa_\m\c^a\pa_\n\c^a+\a\pa_\m\f\pa_\n\f+\frac{1}{d_s}V(\f)g_{\m\n}+2\S_{IJ}(\f)\(F_{\m\r}^I F^J_\n{}^\r-\frac{1}{2d_s}F^I_{\r\s}F^{J\r\s}g_{\m\n}\),\NO\\
&\nabla^\m(Z(\f)\pa_\m\c^a)=0,\qquad 2\a\square\f-V'(\f)=\S'_{IJ}(\f)F^I_{\r\s}F^{J\r\s},\qquad \nabla^\m(\S_{IJ}(\f)F^J_{\m\n})=0.
\eal

\paragraph{Dual frame} As for non-conformal branes \cite{Kanitscheider:2009as}, the holographic dictionary for asymptotically hyperscaling violating Lifshitz backgrounds is best understood in the so called `dual frame' \cite{Chemissany:2014xsa}, obtained by a Weyl rescaling of the Einstein frame metric as  
\be\label{dual-metric}
\lbar g_{\m\n}=e^{-2\x\f}g_{\m\n},
\ee
where the parameter $\x$ is proportional to the hyperscaling violating exponent $\th$ (see eq.~\eqref{xi-theta}).  In the dual frame the action \eqref{action-0} takes the form
\be\label{action}
S_\x=\frac{1}{2\k^2}\int_\cm d^{d_s+2}x\sqrt{-\lbar g}\;e^{d_s\x \f}\Big(\lbar R
-\a_\x (\lbar \pa\f)^2-\S^\x_{IJ}(\f) \lbar F^I_{\m\n}\lbar F^{J\m\n}-Z_\x(\f)(\lbar \pa\c^a)^2-V_\x(\f)\Big)
+S\sbtx{GH}^\x,
\ee 
where we have defined 
\be
\a_\x=\a-d_s(d_s+1)\x^2,\quad
\S^\x_{IJ}(\f)=e^{-2\x\f}\S_{IJ}(\f),\quad
Z_\x(\f)=Z(\f),\quad
V_\x(\f)=e^{2\x\f}V(\f).
\ee
Moreover, the Gibbons-Hawking term in the dual frame is
\be
S\sbtx{GH}^\x=\frac{1}{2\k^2}\int_{\pa\cm} d^{d_s+2}x\sqrt{-\lbar\g}\;e^{d_s\x\f}2\lbar K.
\ee
All quantities with an over-bar are constructed using the dual frame metric \eqref{dual-metric}. In the analysis that follows we will use both the Einstein and dual frame variables, depending on which frame is more convenient for different aspects of the analysis. 

\subsection{Anisotropic black brane solutions}

Depending on the choice of the functions $Z(\f)$, $\S_{IJ}(\f)$ and $V(\f)$, the action \eqref{action-0} admits a wide range of planar solutions, including black holes and domain walls at finite charge density. Such solutions can be asymptotically AdS, conformal to AdS, Lifshitz, or hyperscaling violating Lifshitz. In this paper the functions of the dilaton in the action \eqref{action-0} will be chosen such that the equations of motion admit asymptotically hyperscaling violating Lifshitz backgrounds, with generic values of the Lifshitz and hyperscaling violating exponents $z$ and $\th$.

In the Einstein frame we parameterize the planar background solutions we are interested in as  
\bal\label{Bans}
		&ds^2_B=dr^2+e^{2A(r)}\left(-f(r)dt^2+dx^adx^a\right), \NO\\
		&A_B^I=a^I(r)dt, \qquad \f_B=\f_B(r), \qquad
		\c_B^a=p x^a,
\eal
where again the index $a=1,2,\ldots,d_s$, runs over all spatial dimensions along the conformal boundary and $p\neq 0$ is the isotropic axion charge \cite{Caldarelli:2016nni}. The fieldstrength of the Abelian gauge fields on the backgrounds \eqref{Bans} is given by
\be\label{FB}
F_B^I=\tx dA_B^I=\dot{a}^I\;\tx dr\wedge \tx dt,
\ee
where the dot denotes a derivative with respect to the canonical Einstein frame radial coordinate $r$. In the following we will occasionally use the alternative `domain wall' radial coordinate $u$, defined through the relation  
\be\label{def-u}
\pa_r=-\sqrt{f}\; e^{-A}\pa_u.
\ee
In terms of the radial coordinate $u$ the Einstein frame background metric \eqref{Bans} takes the form
\be\label{dw-metric}
ds^2=e^{2A(u)}\(\frac{du^2}{f(u)}-f(u)dt^2+dx^adx^a\).
\ee

\paragraph{Einstein frame background field equations}
Inserting the ansatz \eqref{Bans} in the field equations \eqref{eoms-0} leads to the system of coupled equations  
\begin{subequations}\label{BEOM}
	\begin{align}
		& d_s\dot A\Big((d_s+1)\dot A+\frac{\dot{f}}{f}\Big)-\a\dot\f^2_B+V(\f_B)+d_sp^2Z(\f_B)e^{-2A}+2f^{-1}e^{-2A}\S_{IJ}(\f_B)\dot a^I\dot a^J=0,\\
		&\ddot{A}+\dot{A}\Big((d_s+1)\dot{A}+\frac{\dot{f}}{2f}\Big)+p^2Z(\f_B)e^{-2A}+\frac{1}{d_s}\Big(V(\f_B)+2e^{-2A}f^{-1}\S_{IJ}(\f_B)\dot{a}^I\dot a^J\Big)=0, \\
		&\ddot{f}+\dot{f}\Big((d_s+1)\dot{A}-\frac{\dot{f}}{2f}\Big)-2p^2fZ(\f_B)e^{-2A}-4e^{-2A}\S_{IJ}(\f_B)\dot{a}^I\dot a^J=0, \\
		& 2\a\ddot{\f}_B+2\a\Big((d_s+1)\dot{A}+\frac{\dot{f}}{2f}\Big)\dot{\f}_B-V'(\f_B)-d_sp^2Z'(\f_B)e^{-2A}+\frac{2}{f}e^{-2A}\S'_{IJ}(\f_B)\dot{a}^I\dot a^J=0,\\
		&\pa_r\left(\S_{IJ}(\f_B) e^{(d_s-1)A}f^{-1/2}\dot{a}^J\right)=0.
	\end{align}
\end{subequations}
The Maxwell equation can be integrated directly to obtain 
\be\label{charge}
\S_{IJ}(\f_B) e^{(d_s-1)A} f^{-1/2}\dot a^J=-q_I,
\ee
where the integration constants $q_I$ are electric charges associated with the Abelian gauge fields $A^I_B$.

\subsection{Asymptotically hyperscaling violating Lifshitz backgrounds}
\label{hvLf-backgrounds}

The system of equations \eqref{BEOM} admits asymptotically hyperscaling violating Lifshitz solutions with generic dynamical exponent $z$ and hyperscaling violation parameter $\th$ provided at least one of the gauge fields, here taken to be $a^1$, has a non-trivial profile, while, at least asymptotically \cite{Charmousis:2010zz,Huijse:2011ef,Iizuka:2011hg,Ogawa:2011bz,Shaghoulian:2011aa,Dong:2012se,Iizuka:2012iv,Gouteraux:2012yr,Gath:2012pg,Chemissany:2014xsa}
\be\label{asym-potentials}
V(\f)\sim-(d_s+z-\th)(d_s+z-\th-1)e^{\frac{2\th}{d_s\m}\f},\qquad \S_{11}(\f)\sim \S_o e^{\frac{2(d_s-1)\th-2d_s^2}{d_s\m}\f},
\ee
where $\S_o$ is an arbitrary positive definite constant and 
\be
\m^2=\frac{(d_s-\th)(d_s z-d_s-\th)}{d_s\a}.
\ee
Recall that the parameter $\a$ corresponds to the normalization of the dilaton kinetic term in \eqref{action-0} and can be specified at will. Moreover, the electric charge of the gauge field supporting the hyperscaling violating Lifshitz background is related to the dynamical exponents as
\be
q_1=\pm\sqrt{\frac{(z-1)(d_s+z-\th)}{32\S_o}}\,.
\ee
The null energy condition (NEC) for this class of solutions requires that $z\geq1$ and $\th\leq d_s+z$.
There exists another class of hyperscaling violating Lifshitz solutions of \eqref{BEOM} with vanishing gauge fields, but those solutions have a fixed hyperscaling violating exponent  $\th=d_s+z$ \cite{Gath:2012pg,Chemissany:2014xsa} and we will not consider them here further. 

In terms of the radial coordinate $r$ in the ansatz \eqref{Bans}, the hyperscaling violating Lifshitz solutions of \eqref{BEOM}  asymptotically take the form
\bal\label{Einstein-background-r}
&ds^2_B\sim dr^2-\(|\th|r/d_s\)^{2-\frac{2d_sz}{\th}}dt^2+\(|\th|r/d_s\)^{2-\frac{2d_s}{\th}}dx^adx^a,\NO\\
&a^1(r)\sim \frac{4{\rm sgn}(\th)q_1}{d_s+z-\th}\(|\th|r/d_s\)^{-\frac{d_s(d_s+z-\th)}{\th}},\quad \f_B(r)\sim -\frac{d_s\m}{\th} \log r,
\eal
while the asymptotic behavior of the second gauge field, $a^2(r)$, depends on the choice of the function $\S_{22}(\f)$ and it is assumed to contribute to the bulk stress tensor at subleading order relative to the gauge field $a^1(r)$ that supports the asymptotic solution. Notice that in these coordinates the UV is located at $r\to\infty$ for $\th<0$ and at $r=0$ for $\th>0$ \cite{Chemissany:2014xsa}. Moreover, the case $\th=0$, corresponding to an asymptotically  Lifshitz background, looks like a singular limit of the asymptotic solution \eqref{Einstein-background-r}.    

A radial coordinate that is better suited for describing hyperscaling violating Lifshitz backgrounds can be defined as \cite{Chemissany:2014xsa}  
\be\label{def-bar-r}
\boxed{d\bar r=-{\rm sgn}(\th)e^{\frac{\th}{d_s\m}\f}dr,}
\ee
which leads to the asymptotic relation 
\be\label{r-bar-r}
r\sim\frac{d_s}{|\th|}e^{-\frac{\th}{d_s}\bar r}.
\ee
In terms of the radial coordinate $\bar r$ the asymptotic solution \eqref{Einstein-background-r} becomes
\bal\label{hvLf-bgnd-asymptotics}
&ds^2_B\sim e^{-\frac{2\th\bar r}{d_s}}\Big(d\bar r^2-e^{2z\bar r}dt^2+e^{2\bar r}dx^adx^a\Big),\NO\\
&a^1(\bar r)\sim \frac{4{\rm sgn}(\th)q_1}{d_s+z-\th}\;e^{(d_s+z-\th)\bar r},\quad \f_B(\bar r)\sim\m \bar r.
\eal
In this coordinate system the metric is manifestly conformally related to a Lifshitz geometry and the solution for $\th=0$ can be smoothly obtained as a limiting case. Moreover, setting the parameter $\x$ that defines the dual frame metric in \eqref{dual-metric} to
\be\label{xi-theta}\boxed{
\x=-\frac{\th}{d_s\m},}
\ee
one can identify the radial coordinate $\bar r$ with the canonical radial coordinate in the dual frame, where the background \eqref{Einstein-background-r} becomes asymptotically Lifshitz, i.e. in the dual frame 
\be\label{dual-metric-background}
d\bar s^2_B\sim d\bar r^2-e^{2z\bar r}dt^2+e^{2\bar r}dx^adx^a.
\ee
Another advantage of the radial coordinate $\bar r$ is that, as follows from the asymptotic relation \eqref{r-bar-r}, the UV is located at $\bar r\to\infty$ for any value of $\th$.

\paragraph{Evading the curvature singularity for $\th>0$} For $\th<0$ the conformal factor in the Einstein frame metric \eqref{hvLf-bgnd-asymptotics} blows up as $\bar r\to \infty$, while it approaches zero for $\th>0$. As a consequence, the Einstein frame metric possesses a well defined conformal boundary for $\th<0$, but for $\th>0$ there is a curvature singularity at $\bar r\to \infty$. In contrast, the dual frame metric \eqref{dual-metric-background} is independent of the hyperscaling exponent $\th$ and, hence, possesses a well defined conformal boundary for all $\th$. Since hyperscaling violating Lifshitz backgrounds necessarily involve a running dilaton that diverges at the UV, it was argued in \cite{Chemissany:2014xsa} that the presence of a metric curvature singularity is a frame-dependent property that can be avoided simply by going to the dual frame. Instead of discarding the case $\th>0$, therefore, we postulate that the holographic dictionary for all values of $\th$ should be constructed in the {\em dual frame}.

\subsection{An exact black brane solution}
\label{example}

An exact asymptotically hyperscaling violating black brane solution of \eqref{BEOM} with non-zero axion charge was presented in \cite{Cremonini:2016avj} (see also \cite{Ge:2016lyn,Bhatnagar:2017twr}) for the case $d_s=2$ and generic exponents $z$ and $\th$, subject to a number of conditions that we will specify shortly. This solution exists when the expressions \eqref{asym-potentials} hold exactly, instead of merely asymptotically near the UV. 
Specifically, the Lagrangian that admits the solutions found in \cite{Cremonini:2016avj} corresponds to the choice of functions
\bal\label{asym-potentials-exact}
&V(\f)=-(2+z-\th)(1+z-\th)e^{\th\f/\m},\qquad \S_{11}(\f)=\frac14 e^{(\th-4)\f/\m},\NO\\
&\S_{12}(\f)=\S_{21}(\f)=0,\qquad \S_{22}(\f)=\frac14 e^{(2z-2-\th)\f/\m},\qquad Z(\f)=\frac12e^{\frac{\m}{\th-2}\f},
\eal
with the normalization of the dilaton kinetic term in \eqref{action-0} chosen such that $\a=1/2$. The exact black brane solution discussed in \cite{Cremonini:2016avj,Bhatnagar:2017twr} then takes the form\footnote{Note the change of notation relative to \cite{Cremonini:2016avj}:
	$r\to v$, $\th\to -\th$, $\g\to\m$, $f\to\cf$, $\a\to p$ and $Q_I\to 4{\rm sgn}(\th) q_I$.}
\bal
\label{HVbackground}
&ds^2 = v^{-\th} \left( - v^{2z} \cf(v) dt^2 + \frac{dv^2}{\cf(v) v^2} + v^2 d\vec{x}^2 \right),\quad \phi = \m \log v, \quad \chi^a = p \, x^a,\NO\\
&a^1= \frac{4{\rm sgn}(\th)q_1}{2+z-\th}(v^{2+z-\th}-v^{2+z-\th}_h),\quad a^2 = \frac{4{\rm sgn}(\th)q_2}{\th-z}(v^{\th-z}-v^{\th-z}_h), 
\eal
where the axion charge $p$ denotes the amount of momentum dissipation in the dual system and
\be
q_1=\pm\sqrt{\frac{(2+z-\th)(z-1)}{8}}\;.
\ee
$v_h$ denotes the location of the horizon and corresponds to the largest real root of the equation $\cf(v)=0$, where the blackening factor is given by
\be\label{blackening-sol}
\cf(v)=1+\frac{p^2}{(2-\th)(z-2)v^{2z-\th}}-\frac{m}{v^{2+z-\th}}+\frac{8q_2^2}{(2-\th)(z-\th)v^{2(z+1-\th)}}.
\ee
Although this expression for the blackening factor holds provided $z\neq 2$, $\th\neq 2$ and $\th\neq z$, the solution can be extended to these cases by including certain logarithmic terms. For example, the blackening factor for the case $z=2$, $\th\neq z$ is given in \cite{Cremonini:2016avj}. The radial coordinate $v$ in \eqref{HVbackground} is related to the radial coordinates, $r$, defined in \eqref{Bans}, $\bar r$, defined in \eqref{def-bar-r}, and $u$ defined in \eqref{def-u} through the identities 
\be\label{coords}\boxed{
	dr=-{\rm sgn}(\th)v^{-\th/2}\cf^{-1/2}(v)\frac{dv}{v},\qquad d\bar r=\cf^{-1/2}(v)\frac{dv}{v},\qquad du={\rm sgn}(\th)v^{z-3}dv.}
\ee
In particular, the coordinate $v$ is asymptotically related to the canonical radial coordinate in the dual frame as 
$v\sim e^{\bar r}$ for all values of $\th$. All of these coordinates are useful for different aspects of the subsequent analysis.

In order for the solution \eqref{HVbackground} to be asymptotically hyperscaling violating Lifshitz with exponents $z$ and $\th$, the blackening factor $\cf(v)$ must asymptote to 1 as $v\to\infty$. For $q_2\neq 0$ and $p\neq 0$, together with the NEC, this requires that $z\geq 1$ and $\th<z+1$. Moreover, the axion charge $p$ is a magnetic charge that corresponds to a non-normalizable mode and should be kept fixed for all solutions of the theory, while the mass $m$ and the electric charge $q_2$ correspond to normalizable modes, i.e. to state-dependent parameters. We therefore demand that the term proportional to $p^2$ is asymptotically dominant relative to the terms involving $m$ and $q_2$, which further restricts the values of the dynamical exponents to
\be\label{exponents}\boxed{
1\leq z <2,\qquad \th<z.}
\ee
Demanding that the exponents $z$ and $\th$ satisfy \eqref{exponents} ensures that the terms in \eqref{blackening-sol} are increasingly asymptotically subleading in the order they are written, from left to right. Note that the upper bounds in these inequalities can be saturated, but we will not consider these cases here since the form of the blackening factor \eqref{blackening-sol} would acquire logarithmic terms, which must be analyzed separately. It is important to emphasize that the conditions \eqref{exponents} are {\em not} a generic requirement for hyperscaling violating backgrounds, but rather a specific feature of the solution \eqref{HVbackground}.

\section{Linear fluctuations and the holographic dictionary}
\label{sec:linear}
\setcounter{equation}{0}

In order to compute the conductivities for hyperscaling violating Lifshitz backgrounds of the form \eqref{Bans} it is necessary to not only solve the system of linearized field equations around such backgrounds, but also to correctly identify the holographic observables, i.e. the sources and the local operators, in the dual theory. As we will see in the subsequent analysis, identifying the holographic observables in such theories is far from trivial and requires a novel form of holographic renormalization.     

In appendix \ref{sec:fluctuations} we derive the linearized equations for a consistent set of spatially homogeneous, time dependent fluctuations around a generic background of the form \eqref{Bans} for the special case $d_s=2$, i.e. four dimensional bulk. The system of equations we derive can be used to compute certain two-point functions at arbitrary frequency and in any background of the form \eqref{Bans}, including asymptotically AdS$_4$ backgrounds. However, in this section we focus exclusively on the zero frequency limit of the fluctuations around the hyperscaling violating Lifshitz background \eqref{HVbackground}.  

Without loss of generality, we consider fluctuations that preserve the Einstein frame gauge
\be
ds^2=dr^2+\g_{ij}(r,t)dx^idx^j,\qquad A_r^I=0,
\ee
where $x^i=\{t,x^a\}$, and we parameterize the fluctuations as
\be\label{fluctuations}
\g_{ij}=\g_{Bij}+h_{ij},\quad 
A_i^I=A_{Bi}^I+\frak a_i^I, \quad
\f=\f_B+\vf,\quad
\c^a=\c_B^a+\t^a.
\ee
Defining $S_i^j\equiv\g_B^{jk}h_{ki}$, the linearized equations allow one to consistently set $S_t^{t}=S_x^x=S_y^y=S_x^y=\vf=\frak a_t=0$ and to consider only the non-zero components $\frak a_a^I=\frak a_a^I(r,t)$, $S_t^a=S_t^a(r,t)$, and $\t^a(r,t)$. The resulting system of linear equations can be reduced to the coupled system of equations \eqref{fluctuation-eqs}, involving only the variables $\frak a^I_a$ and 
\be\label{Theta}
\Th^a\equiv S^a_t-\frac{i\o}{p} \t^a,
\ee
where $\o$ is the frequency. Both $S^a_t$ and $\t^a$ can then be determined from the solution for $\frak a^I_a$ and $\Th^a$.

\subsection{Near horizon solutions and ingoing boundary conditions}

To compute the retarded two-point functions in a generic background of the form 
\eqref{Bans} we need to impose ingoing boundary conditions at the black brane horizon. In the domain wall coordinates \eqref{dw-metric}, near the horizon the blackening factor $f$ takes the form  
\be
f(\r)=4\p T\r+\co(\r^2), \qquad \r\equiv u_h-u,
\ee
where $u_h$ is the smallest root of the equation $f(u)=0$ and $T$ is the Hawking temperature. To leading order in $\r$, the system of linear equations \eqref{fluctuation-eqs} becomes
\begin{subequations}
	\bal
	&\S_{IJ}(\f_B(u_h))\[\r\pa_\r\(\r\pa_\r{\frak a}^J_a\)+\(\frac{\o}{4\p T}\)^2\frak a^J_a\]+2p^2\o^{-2}q_IZ(\f_B(u_h))\r^2\pa_\r \Th^a=0,\\
	&\r\pa_\r\(\r\pa_\r(e^{2A}\Th^a)\)+ \(\frac{\o}{4\p T}\)^2e^{2A}\Th^a-4q_I\r\pa_\r(\r\frak a_a^I)=0.
	\eal
\end{subequations}
It follows that ingoing solutions at the horizon, to leading order in the near horizon expansion, are of the form
\be\label{ingoing}\boxed{
\frak a_a^{I\;(\tx{in})}\propto \r^{-\frac{i\o}{4\p T}},\qquad \Th^{a\;(\tx{in})}\propto e^{-2A}\r^{-\frac{i\o}{4\p T}}.}
\ee

\subsection{Small frequency expansions}

Our next goal is to solve the system of equations \eqref{fluctuation-eqs} for a generic background of the form \eqref{Bans} in a small frequency expansion, i.e. 
\be\label{small-freq-exp}
\frak a^I_a =\frak a_{a}^{I(0)}+\o^2 \frak a_{a}^{I(2)}+\co(\o^4),\qquad \Th^a =\Th^{a(0)}+\o^2 \Th^{a(2)}+\co(\o^4).
\ee
In order to compute the DC conductivities it suffices to determine the $\co(\o^0)$ terms in these expansions only. However, as we will see, identifying the $\co(\o^0)$ solution that corresponds to ingoing boundary conditions at the horizon requires to compute part of the $\co(\o^2)$ terms too.

\paragraph{Solution at $\co(\o^0)$} Inserting the small frequency expansions \eqref{small-freq-exp} in the system of linear equations \eqref{fluctuation-eqs} we find that the leading order terms satisfy 
\begin{subequations}
	\label{fluctuation-eqs-0}
	\bal
	&\pa_r\(\S_{IJ}(\f_B) f^{1/2}e^A\dot{\frak a}^{J(0)}_a-q_I\Th^{a(0)}\)=0,\\
	&\pa_r\(e^{3A}f^{-1/2}\dot\Th^{a(0)}-4q_I\frak a^{I(0)}_a\)-2p^2Z(\f_B) e^Af^{-1/2}\Th^{a(0)}=0.
	\eal
\end{subequations}
The first of these equations can be integrated immediately to obtain
\be\label{1st-integral}
\S_{IJ}(\f_B) f^{1/2}e^A\dot{\frak a}^{J(0)}_a-q_I\Th^{a(0)}=c_I^a(\o),
\ee
where $c_I^a(\o)$ are arbitrary frequency dependent integration constants. Inserting this expression in the second equation in \eqref{fluctuation-eqs-0}, rearranging the resulting equation using the background equations \eqref{BEOM} and \eqref{charge}, and changing to the domain wall coordinates \eqref{dw-metric} we obtain 
\be
\pa_u\(e^{2A}f^{2}\pa_u(f^{-1}\Th^{a(0)})-4c^a_I(\o) a^{I}\)=0.
\ee
Hence, the general solution for $\Th^{a(0)}$ to leading order in the small frequency expansion is
\be\label{theta-sol}\boxed{
	\Th^{a(0)}(u,\o)=\Th_{1}^{a}(\o)f+\Th_{2}^{a}(\o)f\int_{u_*}^u \frac{du'}{e^{2A(u')}f(u')^{2}}+4c^a_I(\o)f\int_{u_*}^u \frac{du'a^I(u')}{e^{2A(u')}f(u')^{2}},}
\ee
where $\Th_{1}^{a}(\o)$, $\Th_{2}^{a}(\o)$ are arbitrary frequency dependent integration constants, and $u_*$ is an unspecified reference point that should be chosen such that the integrals are well defined. Notice that the solution \eqref{theta-sol} is valid for any background of the form \eqref{Bans} and so the choice of the reference point $u_*$ depends on the specific background. Inserting this solution for $\Th^{a(0)}$ in \eqref{1st-integral} determines
\begin{align}
\label{a-sol}
\boxed{
	\begin{aligned}
	\frak a^{I(0)}_a(u,\o)=&\;\frak a^I_{oa}(\o)-\Th_{1}^{a}(\o)a^I(u)-q_J\Th_{2}^{a}(\o)\int_{u_*}^u  du'\S^{IJ}(\f_B(u'))\int_{u_*}^{u'} \frac{du''}{e^{2A(u'')}f(u'')^{2}}\\
	&-\int_{u_*}^u du'\S^{IJ}(\f_B(u')) f^{-1}(u')\(\d^K_J+4q_Jf\int_{u_*}^{u'}\frac{du'' a^K(u'')}{e^{2A(u'')}f(u'')^{2}}\)c^a_K(\o),
	\end{aligned}
}
\end{align}
where $\frak a^I_{oa}(\o)$ are yet another set of arbitrary frequency dependent integration constants. Finally, \eqref{tau} implies that, to leading order in $\o$, $\t^a$ is determined from the equation
\be\label{tau-dot}
\dot\t^{a(0)}=\frac{i\o }{2pZ(\f_B)f}\Big(4e^{-3A}f^{1/2}q_I\frak a^{I(0)}_a-\dot\Th^{a(0)}\Big).
\ee
Recall that the dot $\dot{}$ stands for a derivative with respect to the Einstein frame radial coordinate $r$ defined in \eqref{Bans}.

\paragraph{Ingoing boundary conditions at the horizon} The expressions \eqref{theta-sol} and \eqref{a-sol} correspond to the general solutions of the linearized equations \eqref{fluctuation-eqs} to leading order in the small frequency expansion and suffice for computing the DC conductivities. However, to do so we must first determine the relations among the integration constants imposed by ingoing boundary conditions at the horizon.

In a gauge where the background gauge fields vanish at the horizon, i.e. $a^I(\r)=\co(\r)$ as $\r\to 0$, the near horizon behavior of the solution \eqref{a-sol} is 
\be\label{NH0-2}
\frak a^{I(0)}_a(\r,\o)=  \frak a^I_{oa}(\o)+\Th_{1}^{a}(\o)\co(\r)+\frac{\S^{IJ}(\f_B(u_h))}{4\p T}\(c^a_J(\o)+\frac{q_J\Th_{2}^{a}(\o)}{4\p T e^{2A(u_h)}}\)(\log\r+\co(\r^0)).
\ee
Comparing this with the small $\r$ expansion of the ingoing solution \eqref{ingoing} determines
that for ingoing solutions the integration constants in \eqref{a-sol} must satisfy
\be\label{rel-1}
c_I(\o)+\frac{q_I\Th_{2}^{a}(\o)}{4\p T e^{2A(u_h)}}=-i\o \S_{IJ}(\f_B(u_h))\frak a^J_{oa}(\o).
\ee  
Similarly, the near horizon behavior of the solution \eqref{theta-sol} is
\be\label{NH0-1}
\Th^{a(0)}(\r,\o)= 4\p T\Th_{1}^{a}(\o)\r+\frac{\Th_{2}^{a}(\o)}{4\p T e^{2A(u_h)}}(1+\co(\r\log\r))+c_I^a(\o)\co(\r\log\r).
\ee
However, in order to compare this with the small $\r$ expansion of the ingoing solution \eqref{ingoing} we need to determine $\Th^{a}(\r,\o)$ up to $\co(\o^2)$.

\paragraph{Solution at $\co(\o^2)$} To determine $\Th^{a}(\r,\o)$ to $\co(\o^2)$ we insert the expansions \eqref{small-freq-exp} in \eqref{fluctuation-eqs}, which leads to the $\co(\o^2)$ equations  
\begin{subequations}
	\label{fluctuation-eqs-2}
	\bal
	&\pa_r\(\S_{IJ}(\f_B) f^{1/2}e^A\dot{\frak a}^{J(2)}_a-q_I\Th^{a(2)}\)\\
	&\hskip0.5in+f^{-1/2}e^{-A}\(\S_{IJ}(\f_B)+\frac{2e^{-2A}q_Iq_J}{p^2Z(\f_B)}\) \frak a^{J(0)}_a-\frac{q_I}{2p^2fZ(\f_B)}\dot\Th^{a(0)}=0,\NO\\\NO\\
	&\pa_r\(e^{3A}f^{-1/2}\dot\Th^{a(2)}-4q_I\frak a^{I(2)}_a\)-2p^2Z(\f_B) e^Af^{-1/2}\Th^{a(2)}\NO\\
	&\hskip0.5in+\pa_r\(\frac{1}{2p^2f Z(\f_B)}\(e^{3A}f^{-1/2}\dot\Th^{a(0)}-4q_I\frak a^{I(0)}_a\)\)=0.
	\eal
\end{subequations}
Without loss of generality, the homogeneous solution of these equations can be absorbed into the integration constants $\frak a^I_{oa}(\o)$, $\Th_{1}^{a}(\o)$, $\Th_{2}^{a}(\o)$ and $c^a_I(\o)$ of the  $\co(\o^0)$ solution \eqref{theta-sol} and \eqref{a-sol}. Hence, at $\co(\o^2)$ we are only interested in the inhomogeneous solution of the system \eqref{fluctuation-eqs-2}.

The first equation in \eqref{fluctuation-eqs-2} determines
\be
-\S_{IJ}(\f_B) f\pa_u{\frak a}^{J(2)}_a-q_I\Th^{a(2)}=v_I^a,
\ee
where
\be
v_I^a=\int du  f^{-1}\(\S_{IJ}(\f_B)+\frac{2e^{-2A}q_Iq_J}{p^2Z(\f_B)}\)\frak a^{J(0)}_a+\frac{q_I}{2p^2}\int \frac{du}{fZ(\f_B)}\pa_u\Th^{a(0)}.
\ee
Inserting this in the second equation in \eqref{fluctuation-eqs-2} gives
\be\label{Theta-2-eq}
\pa_u\(e^{2A}f^{2}\pa_u(f^{-1}\Th^{a(2)})\)
=-f\pa_u\(\frac{1}{2p^2f Z(\f_B)}\(e^{2A}\pa_u\Th^{a(0)}+4q_I\frak a^{I(0)}_a\)\)+4\pa_u a^Jv_J^a.
\ee
The near horizon behavior of $\Th^{a(2)}$ can be determined straightforwardly from this equation. Using the near horizon behavior of the $\co(\o^0)$ solution that we obtained above, one can easily deduce that near the horizon $v_I^a=\co(\log\r)^2$. Expanding \eqref{Theta-2-eq} near the horizon then determines 
\be
\Th^{a(2)}= \frac{1}{2p^2Z(\f_B(u_h))4\p T e^{2A(u_h)}}\(-4\p Te^{2A(u_h)}\Th_{1}^{a}(\o)+4q_I\frak a^I_{oa}(\o)\)\log\r+\co(\log\r)^2.
\ee

Combining this with the near horizon expansion \eqref{NH0-1} of the $\co(\o^0)$ solution and comparing with the small $\r$ expansion of the ingoing solution \eqref{ingoing} leads to a second relation among the integration constants, namely
\be\label{IR1}\boxed{
	\Th_{2}^{a}(\o)=\frac{i\o 4\p T}{2p^2Z(\f_B(u_h))}\(-4\p Te^{2A(u_h)}\Th_{1}^{a}(\o)+4q_I\frak a^I_{oa}(\o)\).
}
\ee
Inserting this in the relation \eqref{rel-1} we obtained above gives 
\be\label{IR2}\boxed{
	c^a_I(\o)=-i\o \(\S_{IJ}(\f_B(u_h))+\frac{2q_Iq_J}{p^2Z(\f_B(u_h))e^{2A(u_h)}}\)\frak a^J_{oa}(\o)+\frac{i\o 4\p T q_I}{2p^2Z(\f_B(u_h))}\Th_{1}^{a}(\o).}
\ee
These two relations allow us to express the integration constants $\Th_{2}^{a}(\o)$ and $c^a_I(\o)$ in terms of the arbitrary constants $\Th_{1}^{a}(\o)$ and $\frak a^I_{oa}(\o)$, which we will later identify with sources of certain dual operators. 

The solutions \eqref{theta-sol} and \eqref{a-sol}, together with the identifications \eqref{IR1} and \eqref{IR2}, provide the general solution of the linear equations \eqref{fluctuation-eqs} to lowest order in the frequency and with ingoing boundary conditions at the horizon. These solutions apply to any background of the form \eqref{Bans} and can therefore be used to compute the DC conductivities in a large class of holographic theories with momentum dissipation. However, identifying the operators whose two-point functions these fluctuations compute is a very non-trivial question that requires a careful analysis of the asymptotic behavior of the solutions at the UV, which depends strongly on the specific choice of background.

\subsection{Asymptotic analysis at the UV}

In our analysis of the linear fluctuations so far we have not specified the background other than demanding that it takes the generic form \eqref{Bans}. The asymptotic analysis in the UV, however, and the holographic dictionary, are sensitive to the asymptotic form of the background. We are therefore compelled at this point to make a concrete choice. Ideally, one would like to have general expressions for the DC conductivities valid for any asymptotically hyperscaling violating Lifshitz background described in section \ref{hvLf-backgrounds}. However, even though they are not directly involved in supporting the asymptotic form of the background to leading order, additional matter fields and their asymptotic behavior do in general affect the form of the UV expansions of the fluctuations, and consequently the holographic dictionary. We already encountered an example of the effect extra matter fields can have on the asymptotics in section \ref{example}, where we saw that in order for the exact background \eqref{HVbackground} to have a well defined asymptotic behavior in the UV, the dynamical exponents must lie in the range \eqref{exponents}, which is more stringent than the conditions imposed by the null energy condition.     

In the subsequent analysis, therefore, we consider fluctuations around the specific exact background \eqref{HVbackground}. In particular, we set the functions that specify the bulk Lagrangian as in \eqref{asym-potentials-exact} and we assume that the dynamical exponents satisfy \eqref{exponents}, although we exclude the relativistic case $z=1$, since it requires a separate analysis. Moreover, we use the radial coordinate $v$ for the asymptotic analysis, since this is the natural coordinate for the exact background \eqref{HVbackground} and also it is simply related asymptotically to the canonical radial coordinate $\bar r$ in the dual frame. Crucially, the UV is located at $v\to\infty$ independently of the value of the hyperscaling violation exponent $\th$. In terms of $v$, the functions \eqref{asym-potentials-exact} evaluated on the background \eqref{HVbackground} take the form   
\be\label{model-functions-v}
\S_{11}(\f_B)=\frac14 v^{\th-4},\qquad \S_{22}(\f_B)=\frac14 v^{2z-2-\th},\qquad Z(\f_B)=\frac12v^{-(2z-2-\th)},
\ee
while a comparison between \eqref{HVbackground} and the general background ansatz \eqref{Bans} allows us to read off 
\be
e^{2A}=v^{2-\th},\qquad f=v^{2(z-1)}\cf(v).
\ee

To determine the UV asymptotic expansions of the linear fluctuations \eqref{theta-sol} and \eqref{a-sol} it is useful to introduce the functions 
\be
\J(u)=\int_{u_*}^u \frac{du'}{e^{2A(u')}f(u')^{2}},\qquad \cy^I(u)=\int_{u^*}^u \frac{du'a^I(u')}{e^{2A(u')}f(u')^{2}},
\ee
which in terms of the radial coordinate $v$ become
\bsub
\label{aux}
\bal
\J(v)=&\;{\rm sgn}(\th)\int^v dv\; v^{\th-3z-1}\cf^{-2}(v),\\
\cy^1(v)=&\;\frac{4{\rm sgn}(\th)q_1}{2+z-\th}\Big(-v_h^{2+z-\th}\J(v)+{\rm sgn}(\th)\int^v dv\; v^{-2z+1}\cf^{-2}(v)\Big),\\
\cy^2(v)=&\;\frac{4{\rm sgn}(\th)q_2}{\th-z}\Big(-v_h^{\th-z}\J(v)+{\rm sgn}(\th)\int^v dv\; v^{2\th-4z-1}\cf^{-2}(v)\Big).
\eal
\esub
The small frequency solutions of the fluctuation equations, \eqref{theta-sol} and \eqref{a-sol}, can then be expressed in the $v$ coordinate in the compact form  
\bsub
\label{sols-v}
\bal
\Th^{a(0)}(v,\o)=&\;v^{2(z-1)}\cf(v)\Big(\Th_{1}^{a}(\o)+\Th_{2}^{a}(\o)\J(v)+4c^a_I(\o)\cy^I(v)\Big),\\
\rule{0cm}{1.cm}\frak a^{I(0)}_a(v,\o)=&\;\frak a^I_{oa}(\o)-\Th_{1}^{a}(\o)a^I(v)-{\rm sgn}(\th)q_J\Th_{2}^{a}(\o)\int^v  dv\; \S^{IJ}v^{z-3}\J(v)\NO\\
&-{\rm sgn}(\th)\int^v dv\;\S^{IJ}v^{-(z+1)}\(\cf^{-1}(v)\d^K_J+4q_J v^{2(z-1)}\cy^K(v)\)c^a_K(\o).
\eal
\esub
These expressions, together with \eqref{aux}, allow us to straightforwardly determine the UV asymptotic expansions of the fluctuations.

\paragraph{UV expansions} Using the expression \eqref{blackening-sol} for the function $\cf(v)$, as well as the allowed range \eqref{exponents} for the dynamical exponents, the integral expressions in \eqref{aux} determine that asymptotically
\bsub
\label{uv-exp-aux}
\bal
\J(v)=&\;{\rm sgn}(\th)\Big(\frac{v^{\th-3z}}{\th-3z}+\cdots\Big),\\
\rule{0cm}{1.cm}\cy^1(v)=&\;\frac{-4q_1}{2+z-\th}\Big(\frac{v^{-2(z-1)}}{2(z-1)}+\frac{2p^2v^{\th+2-4z}}{(2-\th)(z-2)(\th-4z+2)}-\frac{(2m-v_h^{2+z-\th})v^{\th-3z}}{\th-3z}+\cdots\Big),\\
\rule{0cm}{1.cm}\cy^2(v)=&\;\frac{4q_2}{\th-z}\Big(-\frac{v_h^{\th-z}v^{\th-3z}}{\th-3z}+\cdots\Big),
\eal
\esub
where the ellipses stand for subleading terms that are not required for the holographic calculation of the conductivities. Inserting these expansions in turn in \eqref{sols-v}, we obtain the UV expansions of the linear fluctuations in the small frequency limit, namely 
\bsub
\label{uv-exp}
\bal
\Th^{a(0)}(v,\o)=&\;v^{2(z-1)}\cf(v)\Bigg(\Th_{1}^{a}(\o)-\frac{8q_1c_1^a(\o)v^{-2(z-1)}}{(z-1)(2+z-\th)}-\frac{32p^2q_1c_1^a(\o)v^{\th+2-4z}}{(2-\th)(z-2)(\th-4z+2)(2+z-\th)}\NO\\
&\hskip-1.7cm+\frac{v^{\th-3z}}{\th-3z}\Big(\frac{16q_1(2m-v_h^{2+z-\th})c_1^a(\o)}{2+z-\th}+\frac{16q_2v_h^{\th-z}c_2^a(\o)}{z-\th}+{\rm sgn}(\th)\Th_{2}^{a}(\o)\Big)+\cdots\Bigg),\\
\rule{0cm}{1.cm}\frak a^{1(0)}_a(v,\o)=&\;\frak a^1_{oa}(\o)-\Th_{1}^{a}(\o)\frac{4{\rm sgn}(\th)q_1(v^{2+z-\th}-v^{2+z-\th}_h)}{2+z-\th}+\frac{2q_1\Th_{2}^{a}(\o)v^{-2(z-1)}}{(z-1)(\th-3z)}\NO\\
&\hskip-1.7cm-\frac{32{\rm sgn}(\th)q_1q_2v_h^{\th-z}v^{-2(z-1)}}{(z-1)(\th-z)(\th-3z)}c_2^a(\o)-4{\rm sgn}(\th)c^a_1(\o)\Bigg(\frac{p^2v^{4-3z}}{(z-2)(\th-4z+2)(4-3z)}\NO\\
&\hskip-1.7cm-\frac{\Big((z+\th-4)m-2(z-1)v_h^{2+z-\th}\Big)v^{-2(z-1)}}{2(z-1)(\th-3z)}\Bigg)+\cdots,\\
\rule{0cm}{1.cm}\frak a^{2(0)}_a(v,\o)=&\;\frak a^2_{oa}(\o)-\Th_{1}^{a}(\o)\frac{4{\rm sgn}(\th)q_2(v^{\th-z}-v^{\th-z}_h)}{\th-z}-\frac{2q_2\Th_{2}^{a}(\o)v^{2(\th-2z)}}{(\th-2z)(\th-3z)}\NO\\
&\hskip-1.7cm-4{\rm sgn}(\th)c^a_2(\o)\Bigg(\frac{v^{2+\th-3z}}{2+\th-3z}-\frac{p^2v^{2\th-5z+2}}{(2-\th)(z-2)(2\th-5z+2)}+\Big(m-\frac{16q_2^2v_h^{\th-z}}{(\th-z)(\th-3z)}\Big)\frac{v^{2(\th-2z)}}{2(\th-2z)}\Bigg)\NO\\
&\hskip-1.7cm+\frac{64{\rm sgn}(\th)q_1q_2}{(2+z-\th)}c_1^a(\o)\Bigg(\frac{v^{2+\th-3z}}{2(z-1)(2+\th-3z)}+\frac{2p^2v^{2\th-5z+2}}{(2-\th)(z-2)(\th-4z+2)(2\th-5z+2)}\NO\\
&\hskip-1.7cm-\frac{(2m-v_h^{2+z-\th})v^{2(\th-2z)}}{2(\th-3z)(\th-2z)}\Bigg)+\cdots.
\eal
\esub
Finally, from these expansions and \eqref{tau-dot} follows that the UV expansion of the axion fluctuations takes the form $\t^{a(0)}(v,\o)=\t_o^a(\o)+\cdots$, where $\t_o^a(\o)$ is an arbitrary integration constant. 
These asymptotic expansions, together with the relations \eqref{IR1} and \eqref{IR2}, encode all information about the thermoelectric DC conductivities. However, in order to extract this information it is necessary to properly identify the local operators and their sources in the dual hyperscaling violating Lifshitz theory, which is the goal of the next subsection.

\subsection{Renormalized holographic observables and Ward identities}
\label{dictionary}

In order to construct the holographic dictionary one must determine a canonical set of symplectic data that parameterize the space of asymptotic solutions \cite{Papadimitriou:2010as}. This symplectic space is then holographically identified with the space of local sources and operators in the dual quantum field theory. As we pointed out in section \ref{hvLf-backgrounds}, a well defined space of asymptotic hyperscaling violating Lifshitz solutions exists for both positive and negative values of the hyperscaling violation exponent $\th$ only in the dual frame, defined by the Weyl transformation \eqref{dual-metric}, where the parameter $\x$ is related to $\th$ as in eq.~\eqref{xi-theta}. Moreover, a natural basis of symplectic variables that can be used to parameterize the space of asymptotic solutions is provided by the set of generalized coordinates and canonical momenta in the radial Hamiltonian formulation of the bulk dynamics, summarized in appendix \ref{ham}.     

Our first goal in this section, therefore, is to determine the asymptotic expansions of all canonical variables in the dual frame in terms of the modes in the solutions of the linear fluctuations equations. Since we do not consider fluctuations of the dilaton, the relation between the linear fluctuations of the fields in the Einstein and dual frames is straightforward. In particular, the asymptotic expansions of the linear fluctuations in \eqref{uv-exp} apply to both the Einstein and dual frames. The relation between the canonical momenta in the two frames is less trivial, but also straightforward. Evaluating the dual frame canonical momenta \eqref{momenta} in the gauge \eqref{dual-FG} and observing that the extrinsic curvature in the dual frame is related to that in the Einstein frame as 
\be
\lbar K_{ij}=\frac{1}{2\lbar N}\dot{\lbar\g}_{ij}=\frac12 e^{\x\f}\pa_r(e^{-2\x\f}\g_{ij})=e^{-\x\f}(K_{ij}-\x\dot\f \g_{ij}),
\ee  
we can express the dual frame canonical momenta in terms of Einstein frame variables as 
\bal\label{momentum-map}
\lbar\p^{ij}=&\;\frac{1}{2\k^2}\sqrt{-\g}\;e^{2\x\f}(K\g^{ij}-K^{ij}),\NO\\
\lbar \p^{i}_I=&\;-\frac{2}{\k^2}\sqrt{-\g}\;\S_{IJ}(\f)
\g^{ij}F^J_{rj},\NO\\
\lbar\p_\f=&\;\frac{1}{\k^2}\sqrt{-\g}\;
(d_s\x K-\a \dot\f),\NO\\
\lbar\p_{\c a}=&\;-\frac{1}{\k^2}\sqrt{-\g}\;Z(\f)\dot\c^a.
\eal

Linearizing these expressions around the background \eqref{Bans} using the fluctuations \eqref{fluctuations} we find that the only non-zero linearized momenta in the dual frame are
\bal\label{lin-momenta}
\lbar\p^{ta[1]}=&\;\frac{1}{4\k^2}e^{2\x\f_B}e^{-(d_s+1)A}f^{-1/2}\pa_r\big(e^{2d_s A}S^a_t\big),\NO\\
\lbar\p^{a[1]}_I=&\;-\frac{2}{\k^2}e^{(d_s-1)A}f^{1/2}\;\S_{IJ}(\f_B)
\(\dot{\frak a}^J_{a}+f^{-1}\dot a^J S^a_t\),\NO\\
\lbar\p^{[1]}_{\c a}=&\;-\frac{1}{\k^2}e^{(d_s+1)A}f^{1/2}Z(\f_B)\dot\t^a,
\eal
where the superscript $[1]$ indicates that these expressions are linear in the fluctuations of the fields. Using the first coordinate transformation in \eqref{coords}, as well as equations \eqref{Theta}, \eqref{tau-dot} and \eqref{model-functions-v}, we can express these linearized momenta in terms of the background \eqref{HVbackground} and the small frequency fluctuations as 
\bal
\lbar\p^{ta[1]}=&\;-\frac{{\rm sgn}(\th)}{4\k^2}v^{\th-z-1}\pa_v\Big(v^{4-2\th}\Big(\Th^{a(0)}+\frac{i\o}{p} \t^{a(0)}\Big)\Big),\NO\\
\lbar\p^{a[1]}_1=&\;\frac{{\rm sgn}(\th)}{2\k^2}
\(v^{z+\th-3}\cf(v)\pa_v{\frak a}^{1(0)}_{a}+4{\rm sgn}(\th)q_1\Big(\Th^{a(0)}+\frac{i\o}{p} \t^{a(0)}\Big)\),\NO\\
\lbar\p^{a[1]}_2=&\;\frac{{\rm sgn}(\th)}{2\k^2}
\(v^{3z-1-\th}\cf(v)\pa_v{\frak a}^{2(0)}_{a}+4{\rm sgn}(\th)q_2 \Big(\Th^{a(0)}+\frac{i\o}{p} \t^{a(0)}\Big)\),\NO\\
\lbar\p^{[1]}_{\c a}=&\;\frac{i\o}{2p\k^2}\Big(-{\rm sgn}(\th)v^{5-z-\th}\pa_v\Th^{a(0)}-4q_I\frak a^{I(0)}_a\Big).
\eal

Finally, inserting the asymptotic expansions \eqref{uv-exp} for the linear fluctuations in these expressions we obtain the asymptotic expansions for the linearized momenta, namely 
\bal\label{uv-exp-momenta}
\lbar\p^{ta[1]}=&\;\frac{{\rm sgn}(\th)}{4\k^2}\Big(-2(z+1-\th)\Th_1^a(\o)v^{z-\th}-\frac{p^2}{z-2}\Th_1^a(\o)v^{-z}+(z-\th)m\Th_1^a(\o)v^{-2}\NO\\
&\hskip-0.9cm+\frac{2(2-\th)}{q_1}\Big(c_1^a(\o)-\frac{i\o}{p}q_1\t_o^a(\o)\Big)v^{2-\th-z}-\frac{(4-2z-\th)p^2 }{(z-2)(\th-4z+2)q_1}c_1^a(\o)v^{2-3z}\NO\\
&\hskip-0.9cm-\Big(\frac{2-z-\th}{\th-3z}\Big)\Big[\frac{8(\th+z-4)mq_1c_1^a(\o)}{(z-1)(2+z-\th)}-\frac{16q_1 v_h^{2+z-\th}c_1^a(\o)}{2+z-\th}-\frac{16q_2v_h^{\th-z}c_2^a(\o)}{\th-z}+{\rm sgn}(\th)\Th_{2}^{a}(\o)\Big]v^{-2z}+\cdots\Big),\NO\\
\lbar\p^{a[1]}_1=&\;-\frac{2}{\k^2}
\Big(c_1^a(\o)-\frac{i\o q_1}{p} \t^{a}_o(\o)+\cdots\Big),\NO\\
\lbar\p^{a[1]}_2=&\;-\frac{2}{\k^2}
\Big(c_2^a(\o)-\frac{i\o q_2}{p} \t^{a}_o(\o)+\cdots\Big),\NO\\
\lbar\p^{[1]}_{\c a}=&\;\frac{i\o{\rm sgn}(\th)}{2p\k^2}\Bigg(\frac{p^2}{z-2}\Th^a_1(\o)v^{2-z}+\frac{p^2c_1^a(\o)}{(4-3z)q_1}v^{4-3z}-4{\rm sgn}(\th)q_1\frak a^1_{oa}(\o)-4{\rm sgn}(\th)q_2\frak a^2_{oa}(\o)\NO\\
&+\Big((z+\th-4)m-2(z-1)v_h^{2+z-\th}-\frac{16q_2^2v_h^{\th-z}}{\th-z}\Big)\Th^a_1(\o)
+\cdots\Bigg).
\eal

The asymptotic expansions \eqref{uv-exp} and \eqref{uv-exp-momenta} provide the symplectic map between the canonical variables in the dual frame, i.e. the fluctuations $\Th^{a(0)}$, $\frak a_a^{I(0)}$, $\t^{a(0)}$ and their conjugate momenta, and the modes $\Th_1^a(\o)$, $\Th_2^a(\o)$, $\frak a_{oa}^I(\o)$, $c_I^a(\o)$ and $\t_o^a(\o)$ parameterizing the solutions of the linearized equations. The covariant fluctuations and their conjugate momenta provide the correct symplectic variables that should be identified with the local sources and operators in the dual theory, but they depend on the radial coordinate, while the sources and operators in the dual theory should be solely expressed in terms of the radially independent modes $\Th_1^a(\o)$, $\Th_2^a(\o)$, $\frak a_{oa}^I(\o)$, $c_I^a(\o)$ and $\t_o^a(\o)$ parameterizing the solutions of the linearized equations. The resolution to this problem is provided by a suitable canonical transformation that diagonalizes the symplectic map between the canonical variables and the modes parameterizing the solutions of the linearized equations. This procedure is a generalization of the method of holographic renormalization that can be applied to a wider class of backgrounds, beyond asymptotically AdS ones \cite{Papadimitriou:2010as}. 

As we will see below, the fluctuations we have turned on contain a non-zero source, namely $\Th_1^a(\o)$, for the energy flux, which is an irrelevant operator in the dual Lifshitz theory \cite{Ross:2011gu}. As a result, the canonical transformation required to renormalize the asymptotically hyperscaling violating theory we consider here is qualitatively different from the usual canonical transformations that implement holographic renormalization in asymptotically AdS backgrounds or asymptotically Lifshitz backgrounds without a source for the energy flux. However, it shares some features with the renormalization of gauge fields in AdS$_2$ and AdS$_3$ \cite{An:2016fzu,Cvetic:2016eiv,Erdmenger:2016jjg}.

\paragraph{Holographic renormalization} In order to determine the canonical transformation that diagonalizes the symplectic map between the canonical variables and the modes parameterizing the solutions of the linearized equations, i.e. the asymptotic expansions \eqref{uv-exp} and \eqref{uv-exp-momenta}, it is necessary to decompose the induced metric $\lbar\g_{ij}$ in the dual frame in time and spatial components in order to account for the non-relativistic scaling. Following \cite{Chemissany:2014xsa} we parameterize $\lbar\g_{ij}$ as  
\be
\lbar\g_{ij}dx^idx^j=-(n^2-n_a n^a)dt^2+2n_adtdx^a+\s_{ab}dx^a dx^b,\qquad a,b=1,\ldots,d_s,
\ee
in terms of the lapse function $n$, the shift function $n^a$ and the spatial metric $\s_{ab}$. These variables are all dynamical since they are components of the induced metric, in contrast to the non-dynamical fields $\lbar N$ and $\lbar N_i$ in the radial decomposition of the bulk metric discussed in appendix \ref{ham}. 

The values of these variables in a generic Einstein frame background of the form \eqref{Bans} are
\be
\s_{ab}^B=e^{-2\x\f_B}e^{2A}\d_{ab},\qquad n_a^B=0,\qquad n_B=e^{-\x\f_B}e^{A}\sqrt{f},
\ee
while the linear fluctuations \eqref{fluctuations} become
\be
e^{-2\x\f_B}h_{tt}=-2n_B\d n,\qquad e^{-2\x\f_B}h_{ta}=e^{-2\x\f_B}h_{at}=\d n_a,\qquad e^{-2\x\f_B} h_{ab}=\d\s_{ab}.
\ee
It follows that, for the components of the linear fluctuations we turn on,
\be
S^a_t=\s_B^{ab}\d n_b,\qquad S^t_a=-\frac{1}{n_B^2}\d n_a,\qquad \d n=0,\qquad \d\s_{ab}=0.
\ee
Using these identities we can express the variational problem of the regularized on-shell action, within the space of fluctuations we consider, in terms of the non-relativistic variables as 
\bal\label{var}
\d S_{\rm reg}=&\;\int d^{d_s+1}x\;\Big(\lbar\p^{ij}\d\lbar\g_{ij}+\lbar\p_{\c\; a}\d\c^a+\lbar\p^i_1\d A_i^1+\lbar\p^i_2\d A_i^2\Big)\NO\\
=&\;\int d^{d_s+1}x\;\Big(2\lbar\p^{ta}\d n_a+\lbar\p_{\c\; a}\d\c^a+\lbar\p^a_1\d A_a^1+\lbar\p^a_2\d A_a^2\Big).
\eal

We now demonstrate that the canonical transformation required to renormalize the symplectic variables within the space of fluctuations we have turned on is generated by the boundary term 	
\bal\label{bt}
S_{\rm b}=&\;-\int d^{d_s+1}x\;A_a^1\lbar\p_1^a\\
&\hskip-0.6cm+\int d^{d_s+1}x\;\Big(h_1(\f)n_a\lbar\p_1^a+h_2(\f)\frac{\lbar\p^a_1
	\lbar\p_{1a}}{\sqrt{-\lbar\g}}+\sqrt{-\lbar\g}\;h_3(\f)n_a\pa_t\c^a+h_4(\f)\lbar\p_{1a}\pa_t\c^a+\sqrt{-\lbar\g}\;h_5(\f)n_an^a\Big),\NO
\eal
where the functions $h_1(\f)$, $h_2(\f)$, $h_3(\f)$, $h_4(\f)$ and $h_5(\f)$ are yet unspecified. Adding this boundary term to the regularized dual frame on-shell action $S_{\rm reg}$ and varying only the field components that get contributions from the fluctuations we turn on here, the variational principle \eqref{var} becomes
\be\label{var-1}
\d(S_{\rm reg}+S_{\rm b})=\int d^{d_s+1}x\;\Big(2\lbar\P^{ta}\d n_a+\lbar\P_{\c\; a}\d\c^a-\bb A_a^1\d\lbar\p^a_1+\lbar\p^a_2\d A_a^2\Big),
\ee
where the canonical variables $\lbar\p^{ta}$, $A_a^1$ and $\lbar\p_{\c a}$ are transformed according to  
\bal\label{transformed-variables}
\lbar\p^{ta}\to&\;\lbar\P^{ta}=\lbar\p^{ta}+\frac12 h_1(\f)\lbar\p^a_1+\frac12\sqrt{-\lbar\g}\;h_3(\f)\pa_t\c^a+\sqrt{-\lbar\g}\;h_5(\f)n^a,\NO\\
A_a^1\to&\;\bb A_a^1=A_a^1-h_1(\f)n_a-2h_2(\f)\frac{\lbar\p_{1a}}{\sqrt{-\lbar\g}}-h_4(\f)\pa_t\c_a,\NO\\
\lbar\p_{\c a}\to&\;\lbar\P_{\c a}=\lbar\p_{\c a}-\sqrt{-\lbar\g}\;\lbar D_t(h_3(\f)n_a)-\lbar D_t(h_4(\f)\lbar\p_{1a}),
\eal
while $A_a^2$ and its conjugate momentum remain unchanged.

Several comments are in order here. Firstly, in writing \eqref{bt} we have only included terms that contribute to the renormalization of the linearized canonical variables in the low frequency limit. In particular, we have not specified the boundary terms required to renormalize the background values of the canonical variables, or terms that involve spatial derivatives or more than one time derivative. Such terms can be determined systematically, but we do not need them here. In fact, this is one of the advantages of formulating holographic renormalization in the language of canonical transformations: it can be carried out directly within the context one works in, without having to first carry out the general procedure and then specialize to the case of interest. 

Secondly, the boundary term \eqref{bt} is gauge invariant and so the renormalized gauge field $\bb A_a^1$ transforms as the unrenormalized gauge potential $A_a^1$. Since the bulk action \eqref{action-0} is also gauge invariant, it follows that the dual theory possesses a $U(1)^2$ global symmetry associated with the bulk gauge fields $A_a^I$.      

Thirdly, notice that the boundary term \eqref{bt} is the generating function of a canonical transformation that is qualitatively different from the standard canonical transformation required for holographically renormalizing asymptotically AdS backgrounds. In particular, the term in the first line of \eqref{bt} is a Legendre transform that changes the boundary condition imposed on the gauge field $A^1_a$ from Dirichlet to Neumann, while the second line in \eqref{bt} is a local function of the canonical momentum $\p^a_1$, which is holographically identified with the source when Neumann boundary conditions are imposed on $A^1_a$. This type of canonical transformation is also required for renormalizing gauge fields in asymptotically AdS$_2$ or AdS$_3$ backgrounds \cite{An:2016fzu,Cvetic:2016eiv,Erdmenger:2016jjg}. As in those cases, the gauge field $A_a^1$ diverges asymptotically in the UV, as is evident from the asymptotic expansion in \eqref{uv-exp}. Recall that the standard counterterms implement a canonical transformation that renormalizes the canonical momenta, while leaving the induced fields unchanged \cite{Papadimitriou:2010as}. However, whenever the electric chemical potential is not the leading term in the asymptotic expansion of a gauge field, the canonical transformation required to render the variational problem well posed renormalizes the gauge potential instead of its conjugate momentum, as for example in \eqref{transformed-variables}. Although the variational principle \eqref{var-1} suggests that Neumann boundary conditions be imposed on the gauge field $A_a^1$, this is not necessarily the case. As we will see shortly, once the symplectic variables have been renormalized by means of the canonical transformation generated by \eqref{bt}, an extra {\em finite} boundary term can be added in order to change the boundary conditions on the gauge field back to Dirichlet, if so desired.
 
Finally, to compare the boundary terms \eqref{bt} with the more standard counterterms for asymptotically Lifshitz theories, e.g. in \cite{Ross:2009ar,Ross:2011gu,Chemissany:2014xsa}, it is necessary to determine the general asymptotic solutions without linearizing around a background, which we leave for future work. Nevertheless, it is important to keep in mind that there are some crucial differences between the system we consider here and typical asymptotically Lifshitz theories. Besides the hyperscaling violation that was considered also in \cite{Chemissany:2014xsa}, the presence of a non-zero axion charge and of the second gauge field affects the asymptotic UV expansions non-trivially, as can be seen from \eqref{uv-exp}. Moreover, in the present analysis we turn on a linear source for the energy flux, which is an irrelevant operator relative to the Lifshitz theory, but only consider massless gauge fields in the bulk.  

In order to determine the functions $h_1(\f)$, $h_2(\f)$, $h_3(\f)$, $h_4(\f)$ and $h_5(\f)$ and demonstrate that the canonical transformation generated by the boundary term \eqref{bt} renormalizes the symplectic variables, we need to evaluate the renormalized canonical variables \eqref{transformed-variables} asymptotically. Linearizing the expressions \eqref{transformed-variables} for the canonically transformed variables using the identities
\be
\sqrt{-\lbar\g}=n_B\sqrt{\s}=v^2n_B=v^{2+z}\cf^{1/2}(v),
\ee
and 
\be
\d n_a=\s_{ab}^B S^b_t=v^2\Big(\Th^{a}+\frac{i\o}{p} \t^{a}\Big),
\ee
we obtain the following expressions for the linearized and renormalized canonical variables:
\bal\label{transformed-variables-explicit}
\lbar\P^{at[1]}=&\;\lbar\p^{at[1]}+\frac12 h_1(\f)\lbar\p^{a[1]}_1+\frac{i\o}{2}h_3(\f)v^{z+2}\cf^{1/2}(v)\t^a+h_5(\f)v^{z+2}\cf^{1/2}(v)\Big(\Th^{a}+\frac{i\o}{p} \t^{a}\Big),\NO\\
\bb a_a^1=&\;\frak a_a^1-h_1(\f)v^2\Big(\Th^{a}+\frac{i\o}{p} \t^{a}\Big)-2h_2(\f)v^{-z}\cf^{-1/2}(v)\lbar\p_{1}^{a[1]}-i\o v^2h_4(\f)\t^a,\NO\\
\lbar\P_{\c a}^{[1]}=&\;\lbar\p_{\c a}^{[1]}-i\o v^{4+z}\cf^{1/2}(v)h_3(\f)\Big(\Th^{a}+\frac{i\o}{p} \t^{a}\Big)-i\o h_4(\f)v^2\lbar\p_{1}^{a[1]}.
\eal

Using the expression \eqref{blackening-sol} for the blackening factor of the background, $\cf(v)$, and the UV expansions \eqref{uv-exp} and \eqref{uv-exp-momenta} for the induced fields and their conjugate momenta, we determine that the transformed variables \eqref{transformed-variables-explicit} are renormalized provided
\bal\label{h-expressions}
&h_1(\f)=-\frac{4q_1{\rm sgn}(\th)e^{(2-z-\th)\f/\m}}{2+z-\th}\Big(1-\frac{ p^2e^{(\th-2z)\f/\m}}{(2-\th)(z-2)}+m e^{-(2+z-\th)\f/\m}\Big),\NO\\\NO\\ 
&h_2(\f)=\left\{\begin{matrix}
	\hskip-2.2cm\frac{\k^2{\rm sgn}(\th)e^{(4-\th)\f/\m}}{2+z-\th}\Big(1-\Big(\frac12+\frac{2-\th}{4-3z}\Big)\frac{ p^2e^{(\th-2z)\f/\m}}{(2-\th)(z-2)}\Big), & z\neq 4/3,\\&\\
	\frac{\k^2{\rm sgn}(\th)e^{(4-\th)\f/\m}}{2+z-\th}\Big(1+\Big(\frac{2-z-\th}{2(2+z-\th)}-(2-\th)\f/\m\Big)\frac{ p^2e^{(\th-2z)\f/\m}}{(2-\th)(z-2)}\Big), & z=4/3,
	\end{matrix}\right.\NO\\\NO\\
&h_3(\f)=\frac{p\;{\rm sgn}(\th)e^{-4z\f/\m}}{2(z-2)\k^2}\Big(1-\frac{(2-z)(z-4+\th)me^{(z-2)\f/\m}}{p^2}\Big),\,\NO\\\NO\\
&h_4(\f)=\left\{\begin{matrix}
	\hskip-2.7cm\frac{p(z-1)\;{\rm sgn}(\th)e^{(2-3z)\f/\m}}{2(z-2)(4-3z)q_1}, & z\neq 4/3,\\&\\
	\frac{p(z-1)\;{\rm sgn}(\th)e^{(2-3z)\f/\m}}{2(z-2)q_1}\Big(\f/\m-\frac{1}{4z-2-\th}\Big), & z=4/3,
	\end{matrix}\right.\NO\\\NO\\
&h_5(\f)=\frac{(z+1-\th){\rm sgn}(\th)}{2\k^2}e^{-(2z+\th)\f/\m}\Big(1-\frac{(3z+1-2\th)p^2e^{(\th-2z)\f/\m}}{2(2-\th)(z-2)(z+1-\th)}\NO\\
&\hskip2.4in+ \frac{(3+2z-2\th)me^{-(2+z-\th)\f/\m}}{2(z+1-\th)}-\frac{12q_2^2e^{-2(z+1-\th)\f/\m}}{(2-\th)(z-\th)}\Big).
\eal
Inserting these expressions for the functions $h_1(\f)$, $h_2(\f)$, $h_3(\f)$, $h_4(\f)$ and $h_5(\f)$ in the UV asymptotic expansions of the transformed variables \eqref{transformed-variables-explicit} gives	
\bal\label{ren-var}
\lbar\P^{at[1]}=&\;-\frac{1}{4\k^2}v^{-2z}\big(\Th_{2}^{a}(\o)+4\m^Ic_I^a(\o)+\cdots\big),\NO\\
\bb a_a^1=&\;\frak a^1_{oa}(\o)-\m^1\Th_{1}^{a}(\o)+\cdots,\NO\\
\lbar\P_{\c a}^{[1]}=&\;-\frac{2i\o}{p\k^2}q_I\bb a^I_{a}(\o)+\cdots,
\eal
where again the ellipses stand for asymptotically subleading terms,
\be\label{chemical-potentials}
\m^1\equiv-\frac{4{\rm sgn}(\th)q_1v^{2+z-\th}_h}{2+z-\th},\qquad \m^2\equiv-\frac{4{\rm sgn}(\th)q_2v^{\th-z}_h}{\th-z},
\ee
are the electric chemical potentials of the background \eqref{HVbackground}, and we have defined 
\be\label{a2}
\bb a_a^2\equiv\frak a^2_{oa}(\o)-\m^2\Th_{1}^{a}(\o),
\ee
which is the leading order term in UV expansion of gauge field $a_a^2$ in \eqref{uv-exp}. From the variational principle \eqref{var-1} follows that these renormalized variables have the correct radial scaling to be the conjugates of respectively $n_a$, $\lbar\p_1^a$ and $\c^a$. Moreover, since they are related to the original symplectic variables by a canonical transformation they are automatically compatible with the symplectic structure of the theory and, therefore, can be directly identified with observables in the dual field theory.

The astute reader will have noticed that the scalar functions \eqref{h-expressions} that define the boundary counterterms surprisingly depend on the mass $m$ and the electric charge $q_2$ of the background, in addition to the axion charge $p$ and the electric charge $q_1$ that is fixed by the Lifshitz boundary conditions. As we have pointed out above, while $p$ and $q_1$ parameterize non-normalizable modes, i.e. boundary conditions, $m$ and $q_2$ correspond to normalizable modes and define a state in the dual field theory. The local counterterms that renormalize a UV complete theory should not depend on the state! However, as we will see shortly, the mode $\Th_1^a$ in the linearized fluctuations corresponds to the source of the energy flux, which is an irrelevant operator in the dual Lifshitz theory \cite{Ross:2011gu}. It is well known that the counterterms required to renormalize a theory perturbed by a source of an irrelevant operator do in fact depend on the background one-point functions, i.e. on the state \cite{vanRees:2011fr}. For example, using \eqref{momentum-map} and \eqref{Bans} one can relate the function $h_5(\f)$ to the background canonical momentum
\be
\lbar\p_{B\; b}^a=-n_B^2\sqrt{-\lbar\g}\;h_5(\f)\d^a_b+\cdots,
\ee
which determines the VEV of the spatial stress tensor in the dual Lifshitz theory. A similar situation was encountered in the computation of two-point functions in a holographic Kondo model \cite{Erdmenger:2016jjg}.   

Finally, notice that the functions $h_2(\f)$ and $h_4(\f)$ contain a logarithmic term, i.e. linear in $\f$, when $z=4/3$. Such terms would normally signify a conformal anomaly, but in the presence of a background running dilaton the notion of a conformal anomaly is ambiguous without reference to a UV fixed point \cite{Chemissany:2014xsa,Taylor:2015glc}. Examples where a conformal anomaly in the presence of a running dilaton can be defined unambiguously include the D4-brane theory with its M5-brane UV completion \cite{Kanitscheider:2009as}, as well as the AdS$_2$ dilaton-gravity theory studied in \cite{Cvetic:2016eiv}, whose UV completions is provided by a two dimensional conformal field theory. However, in the present bottom-up context we are agnostic about the far UV completion of the theory and so we cannot unambiguously define a notion of a conformal anomaly. Nevertheless, it is interesting to observe that the logarithmic terms appear for the unique value of $z$ that, as we will see in section \ref{sec:conductivities}, potentially leads to a linear resistivity at high temperature. 

Having identified the canonical transformation that renormalizes the symplectic variables, we can define different holographic duals by imposing different boundary conditions on the bulk fields through additional {\em finite} boundary terms. In the following we consider the two distinct theories obtained by imposing Dirichlet or Neumann boundary conditions on the spatial components $A_a^1$ of gauge field $A_i^1$, whose time component supports the Lifshitz background.\footnote{We stress that imposing Dirichlet or Neumann boundary conditions on the gauge fields in the bulk {\em is not} equivalent to changing thermodynamic ensemble, as is often stated. Changing boundary conditions in general changes the dual theory \cite{Witten:2003ya}. Moreover, a given theory can be studied in {\em any} thermodynamic ensemble.} Different boundary conditions for scalars and metric fluctuations in Lifshitz backgrounds have been studied in \cite{Andrade:2012xy,Andrade:2013wsa}.

\paragraph{Renormalized observables for Dirichlet boundary conditions} The variational problem \eqref{var-1} in terms of the renormalized canonical variables is well posed provided $\lbar\p^a_1$ is kept fixed, which corresponds to Neumann boundary conditions on the 
gauge field $A_a^1$. In order to impose Dirichlet boundary conditions on $A_a^1$ we need to add the {\em finite} term  
\be\label{D-term}
S_{D}=\int d^{d_s+1}x\;\bb A^1_a\lbar\p^a_1,
\ee
to the renormalized on-shell action and define 
\be\label{Sren}
S^D_{\rm ren}=\lim_{r\to\infty}\big(S_{\rm reg}+S_{\rm b}+S_{D}\big).
\ee
From \eqref{var-1} follows that the variation of the renormalized action \eqref{Sren} takes the form 
\be\label{var-2}
\d(S_{\rm reg}+S_b+S_{D})=\int d^{d_s+1}x\;\Big(2\lbar\P^{ta}\d n_a+\lbar\P_{\c\; a}\d\c^a+\lbar\p^a_1\d\bb A_a^1+\lbar\p^a_2\d A_a^2\Big),
\ee
which is well posed provided $\bb A_a^1$ is kept fixed.

The renormalized variational principle \eqref{var-2} leads to the following identification of local operators and sources in the dual field theory:	
\begin{table}[H]
\begin{center}
	\begin{tabular}{|l|l|l|l|}
			\hline
			&  Operator & Source & Dimension\\\hline\hline
			Energy flux & 	$\ce^a=2\lim_{\lbar r\to\infty}e^{2z\bar r}\lbar\P^{ta}$ & $\Th_1^a=\lim_{\lbar r\to\infty}e^{-2z\bar r}n_a$ & $d_s+2z-\th-1$\\&&&\\
			$U(1)$ currents & $\cj^a_I=\lim_{\lbar r\to\infty}\lbar\p^a_I$  & $\bb a_{oa}^I$ & $d_s+z-\th-1$\\ &&&\\
			Pseudoscalars &  $\cx_a=\lim_{\lbar r\to\infty}\lbar\P_{\c a}$ & $\t_o^a=\lim_{\lbar r\to\infty}\t^a$ & $d_s+z-\th$\\
			\hline
	\end{tabular}	
\end{center}
\caption{Spectrum of operators corresponding to Dirichlet boundary conditions on all fields.}
\label{D-observables}
\end{table}
\vskip-0.3cm\hskip-0.6cm Recall that $\bar r$ is the canonical radial coordinate in the dual frame defined in \eqref{dual-metric-background}. Of course, these observables are only those turned on by the fluctuations we consider here. In particular, the energy density and the spatial stress tensor operators of the energy-momentum complex \cite{Ross:2011gu} are outside the space of fluctuations we turn on. Using the expressions \eqref{ren-var} for the renormalized variables in terms of the modes in the linear fluctuations we obtain
\be\label{operators}\boxed{
\ce^a=-\frac{1}{2\k^2}\big(\Th_{2}^{a}+4\m^Ic_I^a\big),\qquad
\cj_I^a=-\frac{2}{\k^2}
\Big(c_I^a-\frac{i\o q_I}{p} \t^{a}_o\Big),\qquad
\cx_a=-\frac{2i\o}{p\k^2}q_I\bb a^I_{a}.}
\ee
Notice that, as anticipated, the mode $\Th_1^a$ corresponds to the source of the energy flux $\ce^a$, which is an irrelevant operator for $z>1$.

\paragraph{Renormalized observables for Neumann boundary conditions} If the boundary term \eqref{D-term} is not added to the variational principle \eqref{var-1}, then Neumann boundary conditions must be imposed on the gauge field $A_a^1$, i.e. $\lbar\p^a_1$ should be identified with the source of the dual operator and kept fixed. More accurately, since $\lbar\p_1^i$ is constrained by the conservation equation $\pa_i\lbar\p_1^i=0$ it cannot be directly identified with the arbitrary source of a local operator. The general (local) solution of the conservation equation, however, takes the form $\lbar\p_1^i=\frac{2}{\k^2}\lbar\e^{ijk}\pa_j\wt{\bb a}_{ok}$, for some arbitrary  $\wt{\bb a}_{ok}$ that can be identified with the unconstrained source of the dual operator. This is a special case (the $S$-transformation) of the $SL(2,\bb Z)$ transformation on the dual field theory first discussed in \cite{Witten:2003ya}, or equivalently, particle-vortex duality \cite{Herzog:2007ij}. In the bulk it corresponds to electric-magnetic duality. In fact, Neumann boundary conditions on $A_a^1$ are equivalent to Dirichlet boundary conditions for the magnetic dual $\wt A_a^1$. Moreover, had we dualized the gauge field $A_a^1$ in the action \eqref{action-0} from the start, the boundary term \eqref{bt} would take the standard form of the local boundary counterterms and would involve the fieldstrength $\wt F^1_{ij}$ of the magnetic dual gauge field $\wt A_a^1$, in direct analogy to the electric and magnetic frame boundary counterterms computed in \cite{An:2016fzu}.

For the fluctuations we turn on the only non-trivial components of the canonical momentum $\lbar\p_1^i$ are dualized according to  
\be
\lbar\p_1^a=\frac{2i\o}{\k^2}\ve^{ab}\wt{\bb a}_{ob}^1,
\ee
where $\ve^{ab}$ is the Levi-Civita symbol in two dimensions. The spectrum of (dynamical) operators in the theory defined by the variational principle \eqref{var-1} therefore is: 
\begin{table}[H]
\begin{center}
	\begin{tabular}{|l|l|l|l|}
		\hline
		&  Operator & Source & Dimension\\\hline\hline
		Energy flux & 	$\ce^a=2\lim_{\lbar r\to\infty}e^{2z\bar r}\lbar\P^{ta}$ & $\Th_1^a=\lim_{\lbar r\to\infty}e^{-2z\bar r}n_a$ & $d_s+2z-\th-1$\\&&&\\
		$\wt{U(1)}$ current & $\wt\cj_1^a=\frac{2i\o}{\k^2}\ve^{ab}\bb a_{ob}^1$ & $\wt{\bb a}_{oa}^1=\frac{\k^2}{2i\o}\ve_{ab}\lim_{\lbar r\to\infty}\lbar\p^b_1$ & $z+1\,\,\,$ ($d_s=2$)\\ &&&\\
		$U(1)$ current & $\cj^a_2=\lim_{\lbar r\to\infty}\lbar\p^a_2$  & $\bb a_{oa}^2$ & $d_s+z-\th-1$\\ &&&\\
		Pseudoscalars &  $\cx_a=\lim_{\lbar r\to\infty}\lbar\P_{\c a}$ & $\t_o^a=\lim_{\lbar r\to\infty}\t^a$ & $d_s+z-\th$\\
		\hline
	\end{tabular}	
\end{center}
\caption{Spectrum of operators corresponding to Neumann boundary conditions on $\bb A^1_a$.}
\label{N-observables}
\end{table}
\vskip-0.3cm\hskip-0.6cm 
Note that both operators $\wt\cj_1^a$ and $\cj_1^a$ exist in the Dirichlet theory, as well as the Neumann one. However, $\wt\cj_1^a$ is topological in the Dirichlet theory, while $\cj_1^a$ is topological in the Neumann theory. In tables \ref{D-observables} and \ref{N-observables} we list only the dynamical operators, respectively in the Dirichlet and Neumann theories. We should also point out that the particle-vortex dual currents $\wt\cj_1^a$ and $\cj_1^a$ have the same dimension if and only if $\th=0$. 
 
The exchange of the $U(1)$ currents $\cj_1^a$ and $\wt \cj_1^a$ under the change of boundary conditions on the bulk gauge field $A^1$ is effectively a non-perturbative definition of the particle-vortex duality transformation in the dual Lifshitz theory, that applies even in the absence of a Lagrangian description. When the theory admits a Lagrangian description the particle-vortex duality transformation can be constructed explicitly, at least for certain values of the dynamical exponent $z$. For $z=2$ this transformation is reviewed e.g. in appendix A of \cite{Grandi:2012nu}.

\paragraph{Ward identities} The Ward identities in the dual theory are directly related to the first class constraints \eqref{constraints} in the radial Hamiltonian formulation of the theory reviewed in appendix \ref{ham}. The shift symmetry of the axions gives rise to an additional {\em global} Ward identity, which implies that the axion momentum $\lbar\p_{\c a}$ is a total derivative \cite{Caldarelli:2016nni}.   

Within the space of linear fluctuations we consider in this paper the only non-trivial Ward identity comes from the diffeomorphism constraint   
\be\label{WI-constraint}
-2\lbar D_j\lbar\p^{j}_i+\lbar F^{I}_{ij}\lbar\p^j_I+\lbar\p_\f\pa_i\f+\lbar\p_{\c a}\pa_i\c^a=0.
\ee
Linearizing this constraint around a background of the form \eqref{Bans} leads to a single non-trivial condition, namely\footnote{This constraint was also studied in \cite{Andrade:2016tbr} in the context of Lifshitz theories with momentum relaxation. However, the bulk gauge field supporting the Lifshitz asymptotics was massive in that case, which is presumably the reason why the corresponding Ward identity in terms of renormalized variables differs from our \eqref{WID-ren}.}
\be\label{WID}\boxed{
-2\pa_t\lbar\p^{t[1]}_a+\frac{2}{\k^2}q_I\pa_t\frak a^I_a+p\;\lbar\p^{[1]}_{\c a}=0.}
\ee
From the expressions \eqref{lin-momenta} for the linearized momenta we easily see that this constraint is precisely the fluctuation equation \eqref{eqn:Stzdot-fourier-new} in appendix \ref{sec:fluctuations}. 

Using the identities
\bal
\lbar\p^{a[1]}_t=&\;-n_B^2\lbar\p^{ta[1]}+\d n_b\lbar\p_B^{ba}=-v^{2z}\cf(v)\lbar\p^{ta[1]}+\d n_b\lbar\p_B^{ba},\NO\\
\lbar\p^{t[1]}_a=&\;\s^B_{ab}\lbar\p^{bt[1]}+\d n_a\lbar\p_B^{tt}=v^2\lbar\p^{at[1]}+\d n_a\lbar\p_B^{tt},
\eal
and the canonical transformation \eqref{transformed-variables-explicit}, the constraint \eqref{WID} can be expressed in terms of the renormalized variables \eqref{ren-var}. This leads to the Ward identity 
\be\label{WID-ren}\boxed{
\cx_a=-\frac{2i\o}{p\k^2}q_I\bb a^I_{a}(\o),}
\ee
which we already obtained in \eqref{operators}. We therefore see that the correlation functions of the scalar operator dual to the axion fields are entirely determined by the diffeomorphism symmetry.

\section{Thermoelectric Lifshitz DC conductivities}
\label{sec:conductivities}
\setcounter{equation}{0}

Combining the general solution of the fluctuation equations and the identification of the renormalized physical observables in the previous section, we are now in a position to evaluate the renormalized two-point functions for both Dirichlet and Neumann boundary conditions on the gauge field $\bb A_a^1$. As we shall see, the DC conductivities in the Dirichlet and Neumann theories are closely related, but are distinct. This highlights the fact that the boundary conditions at the UV play a crucial role in identifying the physical observables and in computing the conductivities, which cannot be determined in general solely from a near horizon analysis.

\subsection{Two-point functions for Dirichlet boundary conditions}

In terms of the renormalized gauge field modes $\bb a_a^I$, specified in \eqref{ren-var} and \eqref{a2}, the expressions \eqref{IR1} and \eqref{IR2} following from imposing ingoing boundary conditions on the horizon become 
\bal\label{explicit-ops}
\Th_{2}^{a}(\o)
=&\;\frac{i\o 8\p T}{p^2Z(\f_B(u_h))}\Big(q_I \bb a^I_{oa}(\o)+\big(q_I\m^I-\p Te^{2A(u_h)}\big)\Th_{1}^{a}(\o)\Big),\NO\\
c^a_I(\o)=&\;-i\o\(\S_{IJ}(\f_B(u_h))+\frac{2q_Iq_J}{p^2Z(\f_B(u_h))e^{2A(u_h)}}\)\bb a^J_{oa}(\o)\NO\\
&-i\o\Bigg(\frac{2q_I\big(q_J\m^J-\p Te^{2A(u_h)}\big)}{p^2Z(\f_B(u_h))e^{2A(u_h)}}+\S_{IJ}(\f_B(u_h))\m^J\Bigg)\Th_{1}^{a}(\o).
\eal
Inserting these in the renormalized operators of the Dirichlet theory in \eqref{operators} gives
\bal
\ce^a=&\;\frac{2i\o}{\k^2}\Bigg(\frac{2\big(q_I\m^I-\p T e^{2A(u_h)}\big)^2}{p^2Z(\f_B(u_h))e^{2A(u_h)}}+\S_{IJ}(\f_B(u_h))\m^I\m^J\Bigg)\Th_{1}^{a}(\o)\NO\\
&+\frac{2i\o}{\k^2}\Bigg(\frac{2q_J\big(q_I\m^I-\p T e^{2A(u_h)}\big)}{p^2Z(\f_B(u_h))e^{2A(u_h)}}+\S_{IJ}(\f_B(u_h))\m^I\Bigg)\bb a^J_{oa}(\o),\NO\\
\cj^a_I=&\;\frac{2i\o}{\k^2}\Bigg(\frac{2q_I\big(q_J\m^J-\p Te^{2A(u_h)}\big)}{p^2Z(\f_B(u_h))e^{2A(u_h)}}+\S_{IJ}(\f_B(u_h))\m^J\Bigg)\Th_{1}^{a}(\o)\NO\\&+\frac{2i\o}{\k^2}\(\S_{IJ}(\f_B(u_h))+\frac{2q_Iq_J}{p^2Z(\f_B(u_h))e^{2A(u_h)}}\)\bb a^J_{oa}(\o)+\frac{2i\o q_I}{p\k^2} \t^{a}_o(\o),\NO\\
\cx_a=&\;-\frac{2i\o}{p\k^2}q_I\bb a^I_{a},
\eal
from which we read off the two-point functions
\begin{align}
\hskip-0.7cm
\boxed{
\begin{aligned}
\<\cj_I^a(-\o)\cj_J^b(\o)\>=&\;\frac{2i\o}{\k^2}\(\S_{IJ}(\f_B(u_h))+\frac{2q_Iq_J}{p^2Z(\f_B(u_h))e^{2A(u_h)}}\)\d^{ab},\\
\<\ce^a(-\o)\cj_I^b(\o)\>=&\;\<\cj_I^a(-\o)\ce^b(\o)\>
=\frac{2i\o}{\k^2}\Bigg(\frac{2q_I\big(q_J\m^J-\p T e^{2A(u_h)}\big)}{p^2Z(\f_B(u_h))e^{2A(u_h)}}+\S_{IJ}(\f_B(u_h))\m^J\Bigg)\d^{ab},\\
\<\ce^a(-\o)\ce^b(\o)\>=&\;\frac{2i\o}{\k^2}\Bigg(\frac{2\big(q_I\m^I-\p T e^{2A(u_h)}\big)^2}{p^2Z(\f_B(u_h))e^{2A(u_h)}}+\S_{IJ}(\f_B(u_h))\m^I\m^J\Bigg)\d^{ab},\\
\<\cj_I^a(-\o)\cx^b(\o)\>=&\;-\<\cx^a(-\o)\cj_I^b(\o)\>=-\frac{2i\o q_I}{p\k^2}\d^{ab}.
\end{aligned}
}
\end{align}
All other two-point functions are identically zero.

These correlation functions may be simplified by introducing the {\em Lifshitz heat current}\footnote{The heat current can also be defined through the Ward identity obtained from the time component of the first class constraint \eqref{WI-constraint}. This Ward identity involves the energy density $\ce$ and takes the form $\pa_t\ce+\pa_a\cq_D^a=0$ (see e.g. \cite{Hartnoll:2015sea}). However, within the space of linear fluctuations we consider here this identity is satisfied trivially.} 
\be\label{HC-D}\boxed{
\cq^a_D\equiv \ce^a-\m^J\cj^a_J,}
\ee
where $\m^I$ are the electric chemical potentials of the background \eqref{HVbackground} and are given explicitly in \eqref{chemical-potentials}. Notice that in the relativistic limit it reduces to the standard heat current $\ct^{ta}-\m^J\cj^a_J$, in terms of the relativistic stress tensor $\ct^{ij}$. However, for $z>1$, the heat current is an irrelevant operator, as is the energy flux $\ce^a$. Indeed, the scaling dimension of the heat current \eqref{HC-D} is the one expected for a hyperscaling violating Lifshitz theory \cite{Hartnoll:2015sea}. From expressions \eqref{operators} for the renormalized operators in the Dirichlet theory follows that the heat current takes the form
\be
\cq^a_D=-\frac{1}{2\k^2}\Th_{2}^{a}-\frac{2i\o q_I\m^I}{p\k^2}\t^{a}_o,
\ee
and so its correlation functions simplify to
\bal
\<\cq_D^a(-\o)\cj_I^b(\o)\>=&\;-\frac{2i\o}{\k^2}\Bigg(\frac{2q_I\big(\p T e^{2A(u_h)}\big)}{p^2Z(\f_B(u_h))e^{2A(u_h)}}\Bigg)\d^{ab},\NO\\ \<\cq_D^a(-\o)\cq_D^b(\o)\>=&\;\frac{2i\o}{\k^2}\Bigg(\frac{2\big(\p T e^{2A(u_h)}\big)^2}{p^2Z(\f_B(u_h))e^{2A(u_h)}}\Bigg)\d^{ab}.
\eal

From these two-point functions we can read off the thermoelectric DC conductivity matrix  
\be\hskip-0.3cm
\bs_D^{\rm DC}=\left(\begin{matrix}
	T\lbar\bk & T\ba_J \\
	T\ba_I & \bs_{IJ}
	\end{matrix}\right)=\lim_{\o\to 0}\frac{1}{i\o}\left(\begin{matrix}
	\<\cq_D^a(-\o)\cq_D^b(\o)\> & \<\cq_D^a(-\o)\cj_J^b(\o)\> \\&\\
	\<\cj_I^a(-\o)\cq_D^b(\o)\> & \<\cj_I^a(-\o)\cj_J^b(\o)\>
\end{matrix}\right),
\ee
namely
\begin{align}
\label{DC-conductivities-D}\boxed{\boxed{
\begin{aligned}
\lbar\bk^{ab}=&\;\frac{\p sT}{\k^2p^2Z(\f_B(u_h))}\d^{ab},\\
\ba^{ab}_I=&\;-\frac{4q_I\lbar\bk^{ab}}{sT},\\
\bs^{ab}_{IJ}=&\;\frac{2}{\k^2}\S_{IJ}(\f_B(u_h))\d^{ab}+\frac{16q_Iq_J\lbar\bk^{ab}}{s^2T},
\end{aligned}
}}
\end{align}
where $s=4\p e^{2A(u_h)}$ is the entropy density. These conductivities are consistent with the results of \cite{Donos:2014cya,Cremonini:2016avj,Bhatnagar:2017twr}, which were obtained exclusively from a near horizon analysis without identifying the corresponding conserved currents in the dual Lifshitz theory. Notice that the two $U(1)$ currents present in the theory defined by Dirichlet boundary conditions are on the same footing and so the full matrix $\bs_{IJ}^{ab}$ should be identified as a matrix of electric conductivities. 

\subsection{Two-point functions for Neumann boundary conditions}

In order to compute the two-point functions in the theory defined by Neumann boundary conditions we need to invert the expression for the current $\cj_1^a$ in \eqref{explicit-ops} to express $\bb a_{oa}^1$ in terms of the sources in the Neumann theory, namely
\bal
\bb a^1_{oa}(\o)=&\;\(\S_{11}(\f_B(u_h))+\frac{2q_1^2}{p^2Z(\f_B(u_h))e^{2A(u_h)}}\)^{-1}\Bigg[-\frac{2q_1q_2}{p^2Z(\f_B(u_h))e^{2A(u_h)}}\bb a^2_{oa}(\o)\NO\\&-\Bigg(\frac{2q_1\big(q_J\m^J-\p Te^{2A(u_h)}\big)}{p^2Z(\f_B(u_h))e^{2A(u_h)}}+\S_{11}(\f_B(u_h))\m^1\Bigg)\Th_{1}^{a}(\o)+\(\ve^{ab}\wt{\bb a}^1_b-\frac{q_1}{p} \t^{a}_o(\o)\)\Bigg].
\eal
Using this relation in the remaining expressions for the operators in \eqref{explicit-ops} we determine that the non-zero two-point functions in the Neumann case are
\begin{align}
\label{N-2-pt-functions}
\hskip-0.7cm
\boxed{
	\begin{aligned}
	\<\wt\cj^a_1(-\o)\wt\cj^b_1(\o)\>=&\;\frac{2i\o}{\k^2}\ck\d^{ab},\\
	\<\wt\cj^a_1(-\o)\cj_2^b(\o)\>=&\;\<\cj_2^b(-\o)\wt\cj^a_1(\o)\>=\frac{2i\o}{\k^2}\(\frac{2q_1q_2\ck}{p^2Z(\f_B(u_h))e^{2A(u_h)}}\)\ve^{ab},\\
	\<\cj_2^a(-\o)\cj_2^b(\o)\>=&\;\frac{2i\o}{\k^2}\(\S_{22}(\f_B(u_h))+\frac{2q_2^2\cn}{p^2Z(\f_B(u_h))e^{2A(u_h)}}\)\d^{ab},\\
	\<\wt\cj^a_1(-\o)\ce^b(\o)\>=&\;\<\ce^b(-\o)\wt\cj^a_1(\o)\>
	=\frac{2i\o}{\k^2}\(\m^1+\frac{2q_1\ck\big(q_2\m^2-\p Te^{2A(u_h)}\big)}{p^2Z(\f_B(u_h))e^{2A(u_h)}}\)\ve^{ab},\\
	\<\ce^a(-\o)\cj_2^b(\o)\>=&\;\<\cj_2^a(-\o)\ce^b(\o)\>
	=\frac{2i\o}{\k^2}\(\frac{2q_2\cn\big(q_2\m^2-\p T e^{2A(u_h)}\big)}{p^2Z(\f_B(u_h))e^{2A(u_h)}}+\S_{22}(\f_B(u_h))\m^2\)\d^{ab},\\
	\<\ce^a(-\o)\ce^b(\o)\>=&\;\frac{2i\o}{\k^2}\(\frac{2\cn\big(q_2\m^2-\p T e^{2A(u_h)}\big)^2}{p^2Z(\f_B(u_h))e^{2A(u_h)}}+\S_{22}(\f_B(u_h))(\m^2)^2\)\d^{ab},\\
	\<\wt\cj^a_1(-\o)\cx^b(\o)\>=&\;-\<\cx^b(-\o)\wt\cj^a_1(\o)\>=-\frac{2i\o q_1}{p\k^2}\ck\ve^{ab},\\
	\<\cj_2^a(-\o)\cx^b(\o)\>=&\;-\<\cx^a(-\o)\cj_2^b(\o)\>=-\frac{2i\o q_2\cn}{p\k^2}\d^{ab},\\
	\<\cx^a(-\o)\cx^b(\o)\>=&\;\frac{2i\o q_1^2}{p^2\k^2}\ck,
	\end{aligned}
}
\end{align}
where we have defined 
\bal\label{factors}
\ck\equiv&\;\(\S_{11}(\f_B(u_h))+\frac{2q_1^2}{p^2Z(\f_B(u_h))e^{2A(u_h)}}\)^{-1},\NO\\ \cn\equiv&\;\S_{11}(\f_B(u_h))\ck=\(1+\frac{2q_1^2\S^{-1}_{11}(\f_B(u_h))}{p^2Z(\f_B(u_h))e^{2A(u_h)}}\)^{-1}.
\eal
Notice that $\cn$ takes values in the interval $0<\cn<1$.

Introducing the Lifshitz heat current for the Neumann theory\footnote{This form of the heat current in the case of Neumann boundary conditions on $A_a^1$ again follows from the time component of the Ward identity obtained from the constraint \eqref{WI-constraint}. The reason why $\m^1$ does not enter in the heat current for Neumann boundary conditions is that $q_1$ corresponds a to magnetic charge in terms of the dual gauge field $\wt A_a^1$, and $\m^1$ to a magnetic chemical potential. }
\be\boxed{
\cq_N^a\equiv\ce^b-\m^2\cj_2^b,}
\ee
the two-point functions involving the energy flux and the electric current $\cj_2^a$ take the simpler form
\bal
\<\cq_N^a(-\o)\cj_2^b(\o)\>=&\;-\frac{2i\o}{\k^2}\Bigg(\frac{2q_2\cn\big(\p T e^{2A(u_h)}\big)}{p^2Z(\f_B(u_h))e^{2A(u_h)}}\Bigg)\d^{ab},\NO\\ \<\cq_N^a(-\o)\cq_N^b(\o)\>=&\;\frac{2i\o}{\k^2}\Bigg(\frac{2\cn\big(\p T e^{2A(u_h)}\big)^2}{p^2Z(\f_B(u_h))e^{2A(u_h)}}\Bigg)\d^{ab}.
\eal

Reading off the components of the thermoelectric DC conductivity matrix
\be\hskip-0.3cm
\bs_N^{\rm DC}=\left(\begin{matrix}
	T\lbar\bk_N & T\ba_N \\
	T\ba_N & \bs_N
\end{matrix}\right)=\lim_{\o\to 0}\frac{1}{i\o}\left(\begin{matrix}
\<\cq_N^a(-\o)\cq_N^b(\o)\> & \<\cq_N^a(-\o)\cj_2^b(\o)\> \\&\\
\<\cj_2^a(-\o)\cq_N^b(\o)\> & \<\cj_2^a(-\o)\cj_2^b(\o)\>
\end{matrix}\right),
\ee
we obtain
\begin{align}
\label{DC-conductivities-N}\boxed{\boxed{
		\begin{aligned}
		\lbar\bk^{ab}_N=&\;\frac{\p sT\cn}{\k^2p^2Z(\f_B(u_h))}\d^{ab},\\
		\ba_N^{ab}=&\;-\frac{4q_2\lbar\bk_N^{ab}}{sT},\\
		\bs_N^{ab}=&\;\frac{2}{\k^2}\S_{22}(\f_B(u_h))\d^{ab}+\frac{16q_2^2\lbar\bk_N^{ab}}{s^2T}.
		\end{aligned}
}}
\end{align}
Note that the thermal conductivities \eqref{DC-conductivities-N} and \eqref{DC-conductivities-D} satisfy the simple relation $\lbar\bk/\ba = - T s/(4 q)$, as in \cite{Donos:2014cya}.  Moreover, the conductivities \eqref{DC-conductivities-N} are related to those of the Dirichlet theory in \eqref{DC-conductivities-D} by an effective rescaling of the function $Z$ according to $Z(\f_B(u_h))\to Z(\f_B(u_h))/\cn$, where the factor $\cn$ is defined in \eqref{factors}. Nevertheless, the conductivities \eqref{DC-conductivities-N} are not identical with those obtained from the near horizon analysis, which as we have seen above correspond to imposing Dirichlet boundary conditions on $\bb A_a^1$. The boundary conditions at the UV, therefore, crucially affect the computation of the conductivities. The thermoelectric conductivities \eqref{DC-conductivities-N} can alternatively be obtained by dualizing the gauge field $A_i^1$ in the action \eqref{action-0} from the very beginning and imposing Dirichlet boundary conditions on all fields. Presumably, the near horizon analysis of \cite{Donos:2014cya} in the theory with the dual gauge field $\wt A_i^1$ would produce the same result. 

A notable property of the conductivities \eqref{DC-conductivities-N} is that, as their counterparts for Dirichlet boundary conditions in \eqref{DC-conductivities-D}, they obey the  
Wiedemann-Franz law at weak momentum dissipation, i.e. $\lbar {\bk}/\bs\propto T$, at least for low temperatures.
    
Another interesting feature of both electric conductivities \eqref{DC-conductivities-D} and \eqref{DC-conductivities-N} is that they satisfy the lower bound found in \cite{Grozdanov:2015qia}, which in our theory takes the form\footnote{We are grateful to Sa\v{s}o Grozdanov for bringing this bound to our attention. It is unclear whether such a bound exits for the off-diagonal conductivities $\bs^{ab}_{12}$ in the presence of two $U(1)$ gauge fields as in \eqref{DC-conductivities-D}, since \cite{Grozdanov:2015qia} considers only one gauge field.} 
\be\label{bound} 
\bs^{ab}_{11}\geq \frac{2}{\k^2}\S_{11}(\f_B(u_h))\d^{ab},\qquad \bs^{ab}_{22},\,\,\bs^{ab}_N\geq \frac{2}{\k^2}\S_{22}(\f_B(u_h))\d^{ab}.
\ee
This is expected for the conductivities $\bs^{ab}_{11}$ and $\bs^{ab}_{22}$ obtained by imposing Dirichlet boundary conditions on $A_a^1$, since they agree with the near horizon result of \cite{Donos:2014cya}, on which the bound of \cite{Grozdanov:2015qia} relies. More interesting, and reassuring, is the fact that the Neumann conductivity $\bs^{ab}_N$ also satisfies the same bound. A possible explanation for this is that the conductivities \eqref{DC-conductivities-N} can be obtained from a near horizon analysis after dualizing the gauge field $A^1_\m$ in the original action \eqref{action-0}.  
	
A lower bound on the thermal conductivity was also obtained in \cite{Grozdanov:2015djs}. However, in contrast to the bound of \cite{Grozdanov:2015qia} on the electric conductivity, the bound on the thermal conductivity assumes a relativistic conformal UV fixed point, and so it is not directly applicable to the class of theories we study here. Nevertheless, we can see traces of that bound in non-relativistic theories as well. The temperature of the black brane solution \eqref{HVbackground} is given below in eq.~\eqref{Texp}. For $\th\leq 0$, positivity of this temperature requires that\footnote{For $\th>0$ the inequality \eqref{entropy-bound} ceases to hold and so it is unclear whether a lower bound on the thermal conductivity exists in that case.} 
\be\label{entropy-bound}
\frac{\p s}{\k^2p^2Z(\f_B(u_h))}>\frac{16\p^2 v_h^{-\th}}{2\k^2(2-\th)(2+z-\th)},
\ee
which translates to the thermal conductivity bounds
\be\label{bound-thermal}
\frac{\lbar\bk^{ab}}{T}>\frac{16\p^2 v_h^{-\th}}{2\k^2(2-\th)(2+z-\th)}\d^{ab},\qquad  \frac{\lbar\bk^{ab}_N}{T}>\frac{16\p^2 v_h^{-\th}}{2\k^2(2-\th)(2+z-\th)}\cn\d^{ab},
\ee 
respectively for Dirichlet and Neumann boundary conditions on $A_a^1$. These expressions suggest that the thermal conductivity for Dirichlet boundary conditions in non-relativistic theories is still bounded from below, at least for $\th\leq 0$, but the one obtained from Neumann boundary conditions is less constrained, since the factor $\cn$ can be made arbitrarily small, e.g. by taking $p^2$ to be small.

Although the relativistic limit $z\to 1$ does not generically commute with the asymptotic analysis in the UV, naively setting $z=1$ and $\th=0$ one finds that the bounds \eqref{bound-thermal} on the Dirichlet and Neumann conductivities coincide, since for $z=1$, $q_1=0$ and so $\cn=1$, and take the simple form  
\be\label{bound-thermal-relativisitc}
\frac{\lbar\bk^{ab}}{T},\,\,\,\frac{\lbar\bk^{ab}_N}{T} >\frac{8\p^2}{3}\frac{\d^{ab}}{2\k^2}.
\ee 
This lower bound is twice the one found in \cite{Grozdanov:2015djs}, but this is not surprising given that the relativistic conductivities cannot in general be obtained simply by taking the $z\to 1$ limit of the non-relativistic ones. Nevertheless, the result \eqref{bound-thermal-relativisitc} of the naive relativistic limit of the non-relativistic bounds \eqref{bound-thermal} suggests that a lower bound on the thermal conductivity analogous to the one in \cite{Grozdanov:2015djs} exists for non-relativistic theories as well. It would be very interesting to see how universal these bounds are in non-relativistic theories.   
 
Finally, the two-point functions in \eqref{N-2-pt-functions} involving the particle-vortex dual current $\wt \cj_1^a$ give rise to the following thermoelectric conductivities:
\bal\label{magnetic-conductivities}
&\;\lim_{\o\to 0}\frac{1}{i\o}\<\wt\cj_1^a(-\o)\cq_N^b(\o)\>=\frac{2}{\k^2}\(\m^1-\frac{2\p q_1\ck T}{p^2Z(\f_B(u_h))}\)\ve^{ab},\NO\\
&\;\lim_{\o\to 0}\frac{1}{i\o}\<\wt\cj_1^a(-\o)\cj_2^b(\o)\>=\frac{2}{\k^2}\(\frac{8\p q_1q_2\ck}{p^2sZ(\f_B(u_h))}\)\ve^{ab},\NO\\
&\;\lim_{\o\to 0}\frac{1}{i\o}\<\wt\cj_1^a(-\o)\wt\cj_1^b(\o)\>=\frac{2}{\k^2}\ck\d^{ab}.
\eal
In particular, the $\wt\cj_1\wt\cj_1$ conductivity is the inverse of the $\cj_1\cj_1$ one obtained from the Dirichlet theory in \eqref{DC-conductivities-D}, in agreement with particle-vortex duality 
\cite{Herzog:2007ij}. Note that the charge $q_1$ is a background magnetic charge in the Neumann theory, and $\m^1$ is a magnetic chemical potential. Indeed, the thermoelectric conductivities \eqref{magnetic-conductivities} have precisely the expected structure for a background that is purely electric for one gauge field and purely magnetic for the other. This can be checked qualitatively by comparing with the conductivities obtained in \cite{Bhatnagar:2017twr}, where a background magnetic field is turned on for $A_i^2$. Replacing $B^{\rm there}\to q_1$, $q_1^{\rm there}\to q_2$, $q_2^{\rm there}\to 0$, where ``there'' refers to quantities defined in \cite{Bhatnagar:2017twr}, eq.~(49) in \cite{Bhatnagar:2017twr} is analogous to the $\wt\cj_1\wt\cj_1$ conductivity in \eqref{magnetic-conductivities}, eq.~(46) in \cite{Bhatnagar:2017twr} is analogous to the $\wt\cj_1\cj_2$ conductivity in \eqref{magnetic-conductivities}, and eq.~(57) in \cite{Bhatnagar:2017twr} is analogous to the $\wt\cj_1\cq$ conductivity in \eqref{magnetic-conductivities}. This suggests that the thermoelectric DC conductivities on the dyonic background of \cite{Bhatnagar:2017twr} can alternatively be obtained by considering mixed boundary conditions for the gauge field $A_a^2$, but in the purely electric background \eqref{HVbackground}.

\subsection{Temperature dependence and scaling regimes}

Now that we have obtained the general form of the thermoelectric conductivities for Dirichlet and Neumann boundary conditions, we would like to apply these results to the analytical black brane solutions (\ref{HVbackground}) we considered in subsection \ref{example}, and examine the associated temperature dependence.
We are particularly interested in the differences between the possible scaling regimes for the two sets of 
boundary conditions.
We remind the reader that the electrical part of the conductivity matrix for the model of subsection \ref{example} was already  studied in \cite{Cremonini:2016avj}, and therefore we will see some overlap with that discussion.
However, while the analysis of \cite{Cremonini:2016avj} 
provided some preliminary insight into the conductive behavior of the system, 
it only discussed the electric conductivities corresponding to Dirichlet boundary conditions on $\bb A_a^1$ without fully identifying the physical observables in the dual Lifshitz theory, and was therefore incomplete. Moreover, the analysis of \cite{Cremonini:2016avj} considered only the case $\th<0$ ($\th>0$ in the conventions of \cite{Cremonini:2016avj}). Here we revisit those results in light of the analysis of section \ref{sec:linear} and expand on it by examining the full thermoelectric matrix for both Dirichlet and Neumann boundary conditions on $\bb A_a^1$, and for any $\th<z$, both positive and negative.

The temperature $T = -\frac{{\rm sgn}(\th)}{4\pi} v_h^{z+1} \cf^\prime(v_h) $ of the black brane solutions (\ref{HVbackground}) is given by
\be
\label{Texp}\boxed{
	T = -\frac{{\rm sgn}(\th)}{4\p}\Big((z+2-\theta) v_h^z - \frac{8q_2^2}{2-\theta}v_h^{2\theta-z-2} - \frac{p^2}{2-\theta} v_h^{\theta-z}\Big) \, .}
\ee
Note that for the range $1<z<2$, $\theta<z$ we are considering, the first term inside the parenthesis is always positive and the remaining two negative, independently of the sign of $\theta$. 
To identify analytically relatively simple scaling regimes we would like to inspect two different limiting cases, high and low temperatures. These regimes will be sensitive to the sign of $\theta$, 
as we show next:

\begin{itemize}
	\item $\theta<0$ case:\\
	{\bf Large temperatures:} When the hyperscaling violating parameter is negative, the high temperature 
	limit is defined by $ q_2^2 << v_h^{2z-2\theta+2}$ and $ p^2 << v_h^{2z-\theta}$, and (\ref{Texp}) can be well approximated by 
	\be
	\label{largeT}
	T \sim \frac{z+2-\theta}{4\pi} v_h^z \sim q_1^2 v_h^z\, .
	\ee
	This simple expression will facilitate the identification of clean scaling regimes.
	
	{\bf Small temperatures:} On the other hand, the low temperature limit of the theory can be obtained in three ways, either by working with
	\be
	\label{lowT1}
	p^2 v_h^{\theta-z} << q_2^2 v_h^{2\theta-z-2} \lesssim v_h^z \, ,
	\ee
	or alternatively with
	\be
	\label{lowT2}
	q_2^2 v_h^{2\theta-z-2} << p^2 v_h^{\theta-z} \lesssim v_h^z \, ,
	\ee
	or finally by taking the two negative terms in (\ref{Texp}) to be comparable to each other,
	\be
	\label{lowT3}
	q_2^2 v_h^{2\theta-z-2} + p^2 v_h^{\theta-z} \lesssim v_h^z \, .
	\ee
	These conditions ensure that the temperature is small and positive.
	Unfortunately in these cases we don't have a simple analytical expression for $v_h$ as a function of temperature. For generic values of the scaling exponents expression (\ref{Texp}) must be inverted numerically. Still, these regimes will allow us to highlight some of the key differences in the transport behavior associated with Neumann and Dirichlet boundary conditions.
	
	\item
	$\theta>0$ case:\\
	{\bf Large temperatures:}  When the hyperscaling violation parameter is positive 
	the high temperature limit
	corresponds to small values of the horizon radius, and in particular one has either
	\be
	\label{poshighT1}
	T \sim q_2^2 v_h^{2\theta-z-2} \quad \text{when} \quad q_2^2 v_h^{\theta-2} >> p^2 >> q_1^2 v_h^{2z-\theta} \, ,
	\ee
	or alternatively
	\be
	\label{poshighT2}
	T \sim p^2 v_h^{\theta-z} \quad \text{when} \quad p^2 >> q_2^2 v_h^{\theta-2} >> q_1^2 v_h^{2z-\theta} \, .
	\ee
	The case in which the two positive terms in (\ref{Texp}) are of comparable magnitude is also possible, but would lead to similar scalings for $T$ as a function of horizon radius.
	
	{\bf Small temperatures:} Once again, there are several ways to work in the low temperature regime, 
	depending on the hierarchy of scales in the theory.
	One can take
	\be
	\label{poslowT1}
	q_2^2 v_h^{\theta-2} << q_1^2 v_h^{2z-\theta} \lesssim p^2, 
	\ee
	or alternatively 
	\be
	\label{poslowT2}
	p^2 << q_1^2 v_h^{2z-\theta} \lesssim q_2^2 v_h^{\theta-2},
	\ee
	or finally the two positive terms  in (\ref{Texp}) can be of comparable strength, so that
	\be
	\label{poslowT3}
	q_1^2 v_h^{2z-\theta} \lesssim q_2^2 v_h^{\theta-2} + p^2 \, . 
	\ee
\end{itemize}
We can now ask whether in some of these simple regions of parameter space
we can identify clean scaling regimes for the thermoelectric conductivities.

\subsubsection{Dirichlet case}

Starting from the expressions for the thermoelectric conductivities for Dirichlet boundary conditions
\eqref{DC-conductivities-D}, and evaluating them on the black brane background (\ref{HVbackground}), we obtain
\bal
\label{DC-conductivities-D-background}
\lbar\bk^{ab}=&\;\frac{8 \pi^2}{\k^2p^2} \, v_h^{2z-2\theta}\,  T \; \d^{ab},\NO\\
\ba^{ab}_I=&\;-\frac{8\pi q_I}{\k^2p^2} \, v_h^{2z-\theta-2} \; \d^{ab} ,\NO\\
\bs^{ab}_{11}=&\;\frac{1}{2\k^2} \Big(v_h^{\theta-4}+ 16 \frac{q_1^2}{p^2} v_h^{2z-4} \Big) \d^{ab},\NO\\
\bs^{ab}_{12}=&\;\frac{8}{2\k^2}  \frac{q_1 q_2}{p^2} v_h^{2z-4} \d^{ab}   \, ,\NO\\
\bs^{ab}_{22}=&\;\frac{1}{2\k^2} \Big(v_h^{2z-2-\theta}+ 16 \frac{q_2^2}{p^2} v_h^{2z-4}\Big) \d^{ab},
\eal
whose temperature dependence can then be extracted by expressing the horizon radius $v_h$ as a function of $T$.
This is easy to do at large temperatures, but more challenging analytically at small and intermediate values of $T$.
\begin{itemize}
	\item $\theta<0$ case:\\	
	{\bf Large temperatures:} In the regime  where (\ref{largeT}) holds they behave as
	\bal
	\label{DC-conductivities-D-temperature-highT}
	\lbar\bk^{ab} \sim &\;\frac{8 \pi^2}{\k^2p^2} \, T^{\frac{3z-2\theta}{z}} \; \d^{ab},\NO\\
	\ba^{ab}_I \sim &\;-\frac{8\pi q_I}{\k^2p^2} \, T^{\frac{2z-\theta-2}{z}} \; \d^{ab} ,\NO\\
	\bs^{ab}_{11} \sim &\;\frac{8}{\k^2}  \frac{q_1^2}{p^2} T^{\frac{2z-4}{z}} \d^{ab}  ,\NO\\
	\bs^{ab}_{12} \sim &\;\frac{8}{2\k^2}  \frac{q_1 q_2}{p^2} T^{\frac{2z-4}{z}} \d^{ab}  \, ,\NO\\
	\bs^{ab}_{22} \sim &\;\frac{1}{2\k^2} \Big( T^{\frac{2z-2-\theta}{z}}+ 16 \frac{q_2^2}{p^2} T^{\frac{2z-4}{z}} \Big) \d^{ab} \; .
	\eal
	Note that the hierarchy between the two terms in $\bs_{22}$ depends on the magnitude of $q_2, p$ relative to $v_h$. In particular,
	\be
	\label{negcase1}
	\bs_{22} \sim \frac{1}{2\k^2}T^{\frac{2z-2-\theta}{z}} \quad \text{when} \quad q_2^2 << p^2 v_h^{2-\theta},
	\ee
	while instead
	\be
	\label{negcase2}
	\bs_{22} \sim  \frac{8}{\k^2}  \frac{q_2^2}{p^2} T^{\frac{2z-4}{z}} \quad \text{when} \quad p^2 v_h^{2-\theta} << q_2^2 << v_h^{2z-2\theta+2} \, .
	\ee	
	Since $\theta<0$ and $1<z<2$, note that $\bs_{22}$ in  (\ref{negcase1})  can only scale with positive powers of $T$.
	
	The case described by (\ref{negcase2})  is more interesting, since it corresponds to each component of 
	$\bs_{IJ}$ scaling in the same way $\sim T^{\frac{2z-4}{z}}$ as a function of temperature. 
	In particular, we see that the electric conductivities scale as $\sim 1/T$ when $z=4/3$, which would be interesting for systems that may exhibit a 
	linear temperature dependence for the resistivity, $\rho \sim T$. Recall that the $z=4/3$ case was singled out 
	by the analysis of \cite{Hartnoll:2015sea}.

	{\bf Small temperatures:}
	At low temperature it is challenging to obtain analytic expressions for the horizon radius as a function of temperature, which are needed to isolate easily recognizable scaling regimes in temperature. 
	Nonetheless, we can still examine how the thermoelectric matrix scales as a function of $v_h$, 
	in the three limits (\ref{lowT1}), (\ref{lowT2}) and (\ref{lowT3}), to highlight how it differs from the analogous expressions obtained assuming Neumann boundary conditions.
	Notice that the two components $\lbar\bk^{ab}$ and $\ba^{ab}$ in (\ref{DC-conductivities-D-background}) cannot be simplified any further. Thus, 
	we focus on the electric conductivities $\bs^{ab}_{IJ}$. 
	
	When (\ref{lowT1}) holds we find  that 
	\bal
	\label{DC-conductivities-D-background-lowT1}
	\bs^{ab}_{11}=&\;\frac{8}{\k^2} \frac{q_1^2}{p^2} v_h^{2z-4} \d^{ab} \, , \qquad  \bs^{ab}_{12}= \half \frac{q_2}{q_1}\,  \bs^{ab}_{11}\, , \qquad \bs^{ab}_{22}= \frac{q_2^2}{q_1^2} \, \bs^{ab}_{11} \, .
	\eal
	
	Alternatively, in the low temperature range (\ref{lowT2}) we find that the only component of (\ref{DC-conductivities-D-background}) which can be simplified is $\bs^{ab}_{22}$ and takes the form
	\be
	\bs^{ab}_{22}= \frac{1}{2\k^2} v_h^{2z-2-\theta} \d^{ab} \, .
	\ee	
	Finally, when (\ref{lowT3}) holds we have
	\be
	\bs^{ab}_{22} = \frac{1}{2\k^2 p^2} v_h^{4z-2\theta -2}  \d^{ab} \, ,
	\ee
	with the remaining components of the thermoelectric matrix as in (\ref{DC-conductivities-D-background}).
	Thus, we stress that one obtains different low temperature behaviors for the electric conductivities depending on the hierarchy between the different scales in the system, 
	with a variety of scaling regimes possible. While a more extensive analysis would have to be 
	done numerically, these simple estimates provide insight into the richness of this system.

	\item $\theta>0$ case:\\	
	{\bf Large temperatures:} In the regime  (\ref{poshighT1}) the thermoelectric conductivities
	scale as
	\bal
	\label{DC-conductivities-D-temperature-highTpositive}
	\lbar\bk^{ab} \sim &\;\frac{8 \pi^2 q_2^2}{\k^2p^2} \, \left(\frac{T}{q_2^2}\right)^{\frac{z-2}{2\theta-z-2}} \; \d^{ab},\NO\\
	\ba^{ab}_I \sim &\;-\frac{8\pi q_I}{\k^2p^2} \, \left(\frac{T}{q_2^2}\right)^{\frac{2z-\theta-2}{2\theta-z-2}} \; \d^{ab} ,\NO\\
	\bs^{ab}_{11} \sim &\;\frac{1}{2\k^2} \left(\frac{T}{q_2^2}\right)^{\frac{\theta-4}{2\theta-z-2}}
	\d^{ab}, \quad 
	\bs^{ab}_{12} \sim 
	\frac{4}{\k^2}  \frac{q_1 q_2}{p^2} \left(\frac{T}{q_2^2}\right)^{\frac{2z-4}{2\theta-z-2}}
	\d^{ab}, \quad
	\bs^{ab}_{22} \sim 	\frac{2 q_2}{q_1} \,  \bs^{ab}_{12} \, .
	\eal
	It is easy to check that for these ranges of parameters none of the components of the electric matrix
	$\sigma_{IJ}^{ab}$ can scale as $1/T$.
	
	In the regime  (\ref{poshighT2}) on the other hand we have
	\bal
	\label{DC-conductivities-D-temperature-highTpositive2}
	\lbar\bk^{ab} \sim &\;\frac{8 \pi^2 p^2}{\k^2} \, \frac{1}{T}  \; \d^{ab},\NO\\
	\ba^{ab}_I \sim &\;-\frac{8\pi q_I}{\k^2p^2} \, \left(\frac{T}{p^2}\right)^{\frac{2z-\theta-2}{\theta-z}} \; \d^{ab} ,\NO\\
	\bs^{ab}_{11} \sim &\;\frac{1}{2\k^2} \left(\frac{T}{p^2}\right)^{\frac{\theta-4}{\theta-z}}
	\d^{ab}, \quad 
	\bs^{ab}_{12} \sim 
	\frac{4}{\k^2}  \frac{q_1 q_2}{p^2} \left(\frac{T}{p^2}\right)^{\frac{2z-4}{\theta-z}}
	\d^{ab}, \quad
	\bs^{ab}_{22} \sim 	\frac{1}{2\k^2}  \left(\frac{T}{p^2}\right)^{\frac{2z-2-\theta}{\theta-z}}
	\d^{ab} \, .
	\eal
	As for the case above, none of the electric conductivities here can scale as $1/T$, for our range of $z$ and 
	$\theta$.

	{\bf Small temperatures:} 
	In the low temperature regime, the only case for which the thermoelectric matrix simplifies significantly
	corresponds to the case (\ref{poslowT2}). One can then show that the electrical conductivities scale as they do in   (\ref{DC-conductivities-D-background-lowT1}), which we recall 
	corresponds to $\theta<0$.
	
\end{itemize}

\subsubsection{Neumann case}
Recall that the thermoelectric conductivities in the case of Neumann boundary conditions are modified by a factor of 
\be
\cn = \Big(1+ 16 \frac{q_1^2}{p^2} v_h^{2z-\theta} \Big)^{-1} \, ,
\ee
as compared to those associated with Dirichlet boundary conditions.
Thus, after evaluating on the background we have
\bal
\label{DC-conductivities-N-background}
\lbar\bk^{ab}_N=&\;\frac{8 \pi^2}{\k^2p^2} \, v_h^{2z-2\theta}\,  T \,  \Big( 1+ 16 \frac{q_1^2}{p^2} v_h^{2z-\theta} \Big)^{-1} \; \d^{ab},\NO\\
\ba^{ab}_N=&\;-\frac{8\pi q_2}{\k^2p^2} \, v_h^{2z-\theta-2} \,  \Big( 1+ 16 \frac{q_1^2}{p^2} v_h^{2z-\theta} \Big)^{-1}  \; \d^{ab} ,\NO\\
\bs^{ab}_{N}=&\;\frac{1}{2\k^2} \Big[  v_h^{2z-2-\theta}+ 16 \frac{q_2^2}{p^2} v_h^{2z-4}  \Big( 1+ 16 \frac{q_1^2}{p^2} v_h^{2z-\theta} \Big)^{-1}  \Big]\d^{ab}.
\eal
For generic values of $v_h$ these differ significantly from their Dirichlet counterparts.
Again, we consider separately positive and negative values of the hyperscaling violating exponent:
\begin{itemize}
	\item  $\theta<0$ case:\\
	{\bf Large temperatures:} 
	In the regime (\ref{largeT}) it can be easily shown that $\cn$ scales like
	\be
	\label{cnlimit}
	\cn \sim \frac{p^2}{16 q_1^2} v_h^{\theta-2z} \sim  \frac{p^2}{16 q_1^2} T^{\frac{\theta-2z}{z}} \, ,
	\ee
	and therefore the leading behavior of the components of the thermoelectric matrix is
	\bal
	\label{DC-conductivities-N-background-temperature-highT}
	\lbar\bk^{ab}_N \sim &\;  \frac{ \pi^2}{2 \k^2  q_1^2} \, T^{\frac{z-\theta}{z}}\d^{ab},  \NO\\
	\ba^{ab}_N\sim&\;-\frac{ \pi q_2}{2 \k^2 q_1^2} \, T^{-\frac{2}{z}} \; \d^{ab} ,\NO\\
	\bs^{ab}_{N}\sim&\;\frac{1}{2\k^2} T^{\frac{2z-2-\theta}{z}} \;\d^{ab} .
	\eal
	Notice that while the temperature behavior of $ \lbar\bk^{ab}_N $ and $ \ba^{ab}_N$ is markedly different from their Dirichlet counterparts 
	$ \lbar\bk^{ab} $ and $ \ba^{ab}_2$ in (\ref{DC-conductivities-D-temperature-highT}),
	the electric conductivity $\bs^{ab}_{N} $ in the high temperature limit agrees with the first term in 
	$\bs_{22}^{ab}$. Thus, the two scale in the same way with temperature in the regime $ q_2^2 << p^2 v_h^{2-\theta}$, as 
	seen in (\ref{negcase1}). On the other hand, in the regime defined by  (\ref{negcase2}) the behavior of 
	$\bs^{ab}_{N} $ differs from that of $\bs^{ab}_{22} $. 
	
	Finally, notice that for the range $1<z<2$, $\theta<0$ of scaling exponents we are working with in this regime, 
	$\bs_N$ can only scale as a positive power of $T$. As a result, a resistivity which is linear in temperature (or even quadratic) is not allowed in this case, at least not to leading order. This compels us to examine 
	the low temperature regime.

	{\bf Small temperatures:} 	In the low temperature regime (\ref{lowT1}) the quantity $\cn$ is still given by (\ref{cnlimit}), and we find  the following simple scalings as a function of horizon radius,
	\bal
	\label{DC-conductivities-N-background-lowT1}
	\lbar\bk^{ab}_N=&\;\frac{ \pi^2}{2\k^2 q_1^2} \, v_h^{-\theta}\,  T \; \d^{ab},\NO\\
	\ba^{ab}_N=&\;-\frac{\pi q_2}{2\k^2 q_1^2} \, v_h^{-2}   \; \d^{ab} ,\NO\\
	\bs^{ab}_{N}=&\;\frac{1}{2\k^2}  v_h^{2z-2-\theta}\d^{ab}.
	\eal
	While the first two coefficients differ substantially from the analogous expressions in (\ref{DC-conductivities-D-background}), the electrical 
	conductivity $\bs^{ab}_{N}$ is identical to the first term in $\bs^{ab}_{22}$.
	
	On the other hand, in the low temperature ranges (\ref{lowT2}) and  (\ref{lowT3}) the quantity $\cn$ cannot be simplified further, and thus we have to resort to 
	using the full expression (\ref{DC-conductivities-N-background}), whose structure is highly non-trivial.
	Identifying clean scaling regimes analytically for the electrical conductivity in this case would be challenging
	and one should resort to numerics.
	
	\item $\theta>0$ case:
	
	{\bf Large temperatures:} 
	In the high temperature regimes (\ref{poshighT1}) and (\ref{poshighT2}) the rescaling factor $\cn \sim 1$ and therefore 
	the thermoelectric matrix agrees with the corresponding expressions (\ref{DC-conductivities-D-temperature-highTpositive} and (\ref{DC-conductivities-D-temperature-highTpositive2}) for Dirichlet boundary conditions, if we identify $\lbar\bk^{ab}_N = \lbar\bk^{ab}, \ba^{ab}_N=\ba^{ab}_2, \bs^{ab}_{N} = \bs^{ab}_{22}$.
	
	{\bf Small temperatures:} 
	At low temperature the thermoelectric matrix simplifies significantly only when
	(\ref{poslowT2}) is satisfied, for which we have
	\be
	\cn \sim \frac{p^2}{16 q_1^2} v_h^{\theta-2z} \, .
	\ee
	In this case we find that (\ref{DC-conductivities-N-background-lowT1}) still describes the 
	thermoelectric matrix, which we stress once more is very different from its Dirichlet counterpart.
\end{itemize}

To summarize, at large temperatures the difference between the two sets of boundary conditions is most apparent for the case $\theta<0$. 
When $\theta>0$ and $T$ is large, we find that $\cn \sim 1$ and thus the components of the thermoelectric matrix for Neumann boundary conditions scale in the same way as the corresponding Dirichlet components.

For a negative value of the hyperscaling violating exponent, on the other hand, the two sets of boundary 
conditions yield very different results. 
The only partial agreement is for the special case of (\ref{negcase1}), in which the Neumann electric conductivity
$ \bs_{N}$ behaves in the same way with temperature as its Dirichlet counterpart $\bs_{22}$.

As mentioned earlier, because of the complexity of the background solution we work with, the only regime in which we can 
analytically estimate the temperature scaling of the thermoelectric matrix is that of large temperatures.
As we already mentioned, an interesting phenomenological question is whether one can obtain a robust explanation for the linear scaling of the resistivity $\rho \sim T$ for the strange metal phase.
In this respect a particularly interesting case is described by the regime (\ref{negcase2}) for $\theta<0$ and 
Dirichlet boundary conditions. 
In this case all the components of the electrical DC conductivity $\bs_{IJ}$ scale in the same way $\sim T^{(2z-4)/z}$ and in 
particular for $z=4/3$ scale as $1/T$, leading to a linear resistivity.
As it turns out, in our analysis this is the only case which could in principle support a linear resistivity, 
in the high temperature regime (independently of choices of boundary conditions).
This is intriguing because the special value $z=4/3$ was the one seemingly needed by the scaling arguments and field theoretic analysis of \cite{Hartnoll:2015sea}.
To better understand the significance of this point we would like to generalize this analysis to include a magnetic field, and extend the computation of observables to the Hall angle and the magnetoresistance. This would also complement the recent analysis of \cite{Cremonini:2017qwq,Blauvelt:2017koq}, in which the special values $z=4/3$, $\theta =0$ singled out by \cite{Hartnoll:2015sea} were associated with a minimal DBI holographic model which could be used to reproduce, in the probe limit, the scalings of the resistivity $\rho \sim T$ and 
Hall angle $\cot \Theta_H \sim T^2$ of the cuprates \cite{Blauvelt:2017koq}.   

Finally, at arbitrary and low temperatures the expressions for the thermoelectric conductivities in the Neumann and Dirichlet cases generically differ from each other -- for both positive and negative values of $\theta$ --  and a more extensive analysis of their temperature 
behavior would have to be done numerically.

\section{Concluding remarks}
\label{sec:conclusion}
\setcounter{equation}{0}

In order to holographically identify the conserved currents responsible for the thermoelectric conductivities in a hyperscaling violating Lifshitz theory, it is necessary to place it at the UV, i.e. consider bulk solutions that are asymptotically locally hyperscaling violating Lifshitz, and to construct the physical observables through holographic renormalization. In this paper we have carried out this procedure for linearized fluctuations of the theory \eqref{action-0} around the family of hyperscaling violating Lifshitz backgrounds \eqref{HVbackground}. This analysis was considerably more involved compared to that for relativistic theories because the heat current involves the energy flux, which is an irrelevant operator in the dual Lifshitz theory when $z>1$. In particular, we showed that the boundary counterterms required to renormalize the theory in the presence of a source for the energy flux involve the radial canonical momentum conjugate to the gauge field supporting the Lifshitz asymptotics, in close resemblance to the renormalization of gauge fields in asymptotically AdS$_2$ and AdS$_3$ backgrounds \cite{Cvetic:2016eiv,Erdmenger:2016jjg}. However, if the gauge field supporting the Lifshitz asymptotics in the action \eqref{action-0} is dualized to its magnetic dual, the boundary counterterms would assume a more standard form that does not involve the canonical momenta  \cite{An:2016fzu}.

An additional complication in the identification of the dual operators responsible for the thermoelectric conductivities is related to the choice of possible boundary conditions on the bulk fields. It is well known that the field theory dual of a given gravitational theory in the bulk is fully specified only once boundary conditions at infinity are imposed. In particular, the spectrum of local operators and their correlation functions are sensitive to the boundary conditions. Holographic conductivities are no exception. In this paper we have explicitly demonstrated the dependence of physical observables on the boundary conditions by computing analytically the thermoelectric DC conductivity matrix in a hyperscaling violating Lifshitz theory with two different boundary conditions on one of the two Maxwell fields present. Only Dirichlet boundary conditions lead to conductivities that agree with the result of the near horizon analysis \cite{Donos:2014cya}, while Neumann boundary conditions produce a different set of DC conductivities. However, we expect that a near horizon analysis in the theory obtained by dualizing the Maxwell field supporting the Lifshitz background should reproduce the thermoelectric conductivities obtained by imposing Neumann boundary conditions in the original theory. It would be interesting to confirm this explicitly.  

Although the thermoelectric conductivities we computed have a rich behavior as functions of the temperature, depending on the various parameters characterizing the background solution, we carried out a preliminary analysis by looking for parameter regions where the DC conductivities exhibit approximate scaling behavior with the temperature. We identified several clean scaling regimes in the limit of large temperature, which in our setup probes the Lifshitz theory in the UV. The thermoelectric conductivities obtained from Dirichlet and Neumann boundary conditions on the Maxwell field supporting the Lifshitz background scale generically differently with temperature. Rather intriguingly, the only case we could identify that can potentially lead to a linear resistivity arises in the case of Dirichlet boundary conditions and the specific value of the dynamical exponent $z=4/3$, which was also singled out in the field theory analysis of \cite{Hartnoll:2015sea}.        

The work presented in this paper can be extended in several promising directions. An interesting generalization is to compute the optical (AC) thermoelectric conductivities for the same model and background we consider here. This should be relatively straightforward since the linearized equations we have obtained in appendix \ref{sec:fluctuations} hold for arbitrary frequency, but solving them for general frequency would require extensive numerical analysis. In particular, one can ask what kinds of clean scaling laws can be supported in the intermediate frequency regime, along the lines of the analysis done in \cite{Bhattacharya:2014dea}.

Another obvious extension of the current analysis is to include a background magnetic field, as was done for the model we study here in \cite{Bhatnagar:2017twr}. A much simpler way to obtain the thermoelectric DC conductivities in a dyonic background, however, is to solve the linearized fluctuation equations for a purely electric background as we have done in this paper, and simply impose mixed boundary conditions on the gauge field $A_a^2$. Both procedures are equivalent and place the dual field theory on a dyonic background. In particular, the presence of a background magnetic field will facilitate the evaluation of the Hall angle and the magnetoresistance in the dual Lifshitz theory. These in turn will permit a more detailed comparison of the holographic DC conductivities with experimental observations for the cuprates. 

Yet another potential direction to explore is turning on a mass for the bulk gauge field supporting the Lifshitz asymptotics so that the dual Lifshitz theory is characterized by an additional vector exponent \cite{Gouteraux:2012yr}. This again would allow for a more direct comparison with the field theory results of \cite{Hartnoll:2015sea}. Finally, it would be very interesting to revisit part of the asymptotic analysis of \cite{Chemissany:2014xsa} for generic asymptotically locally hyperscaling violating Lifshitz theories in order to include a non-zero source for the energy flux and consider the effect of different boundary conditions on the bulk gauge field. We hope to address some of these questions in the near future.

\section*{Acknowledgments}

S.C. would like to thank Blaise Gouteraux, Richard Davison and Li Li for valuable discussions.
This research is supported in part by the DOE Grant Award DE-SC0013528, (M.C.), the Fay R. and Eugene L. Langberg Endowed Chair (M.C.) and the Slovenian Research Agency (ARRS) (M.C.). 
The work of S.C. is supported in part by the National Science Foundation grant PHY-1620169.
M.C. thanks CERN, and I.P. thanks CERN and the University of Pennsylvania, for the hospitality during the completion of this work.

\appendix

\renewcommand{\thesection}{\Alph{section}}
\renewcommand{\theequation}{\Alph{section}.\arabic{equation}}

\section*{Appendices}
\setcounter{section}{0}

\section{Fluctuation equations in the Einstein frame}
\label{sec:fluctuations}
\setcounter{equation}{0}

In this appendix we derive the Einstein frame linearized field equations for a consistent set of spatially homogeneous, {\em time dependent} fluctuations around a generic background of the form \eqref{Bans} for the case $d_s=2$, i.e. four dimensional bulk. In particular, the system of equations we derive can be used to compute the AC conductivities too, but in this paper we only solve them in the zero frequency limit. Moreover, we keep the background completely arbitrary and so the same linearized equations can be used to compute two-point functions in any background of the form \eqref{Bans}. The analysis of the linearized fluctuations in this appendix is almost identical to that in \cite{Lindgren:2015lia}, except that here we turn a background axion charge, but we set the background magnetic field to zero.  

We choose to work in a gauge where the Einstein frame metric takes the form
\be
ds^2=dr^2+\g_{ij}(r,t)dx^idx^j,
\ee
where $i,j$ run over the time $t$ and spatial dimensions $x^a$, and the radial component of the gauge fields is set to zero, i.e. $A^I_r=0$. Parameterizing the most general fluctuations that preserve this gauge by
\be
\g_{ij}=\g_{Bij}+h_{ij},\quad 
A_i^I=A_{Bi}^I+\frak a_i^I, \quad
\f=\f_B+\vf,\quad
\c^a=\c_B^a+\t^a,
\ee
with $S_i^j\equiv\g_B^{jk}h_{ki}$,  we turn off the fluctuation components $S_t^{t}=S_x^x=S_y^y=S_x^y=\vf=\frak a_t=0$ and only keep the components $\frak a_a^I=\frak a_a^I(r,t)$, $S_t^a=S_t^a(r,t)$, and $\t^a(r,t)$. Inserting these fluctuations in the Einstein frame equations \eqref{eoms-0} leads to the following set of linear equations:
\begin{subequations} \label{eqn:Fluctuations}
	\begin{align}
	&\mbox{\bf Einstein $ta$:}\NO\\
	&\Big(\pa_r^2+\Big(3\dot{A}-\frac{\dot f}{2f}\Big)\pa_r-2p^2Z(\f_B)e^{-2A}\Big)S_t^a=-2e^{-2A}\(pZ(\f_B)\pa_t\t^a+2\S_{IJ}(\f_B)\dot{a}^I\dot{\frak a}_a^J\),\label{eqn:Stx} \\
	&\Big(\pa_r^2+3\Big(\dot{A}+\frac{\dot f}{2f}\Big)\pa_r+\frac{4}{f} e^{-2A}\S_{IJ}(\f_B)\dot a^I\dot a^J\Big)S_a^t=\frac{2}{f}e^{-2A}\(pZ(\f_B)\pa_t\t^a+2\S_{IJ}(\f_B)\dot{a}^I\dot{\frak a}_a^J\),\label{eqn:Stx-reverse} \\
	&\mbox{\bf Einstein $ra$:}\NO\\
	&\pa_t\dot S^a_t=-4e^{-2A}\S_{IJ}(\f_B)\dot a^I\pa_t \frak a_a^J-2pZ(\f_B)f\dot\t^a,\label{eqn:Stxdot} \\
	&\mbox{\bf Maxwell $a$:}\NO\\
	&\pa_r\left(\S_{IJ}(\f_B) f^{-1/2}e^A\left(\dot a^J S^a_t+f\dot{\frak a}_a^J\right)\right)=\S_{IJ}(\f_B) f^{-1/2}e^{-A}\pa_t^2\frak a_a^J,\label{eqn:Maxx}\\
	&\mbox{\bf Axion:}\NO\\
	&\ddot\t^a+\left(3\dot A+\frac12f^{-1}\dot f+Z^{-1}(\f_B)Z'(\f_B)\dot\f_B\right)\dot\t^a-f^{-1}e^{-2A}\pa_t^2\t^a=pe^{-2A}\pa_tS^t_a,\label{eqn:axion}
	\end{align}
\end{subequations}
and we recall that $\dot{}\equiv\pa_r$. Notice that the first two equations, (\ref{eqn:Stx}) and (\ref{eqn:Stx-reverse}), are in fact not independent since
\be
S^t_a=-f^{-1}S^a_t.
\ee

Using the Maxwell equation for the background, eq.~\eqref{charge}, and Fourier transforming in time ($\pa_t\to i\o$) we can write these equations in the form 
\begin{subequations} \label{eqn:Fluctuations-fourier-new}
	\begin{align}
	&\pa_r\left(e^{3A}f^{-1/2}\dot S^a_t \right)=4q_I\dot{\frak a}^I_a+2pZ(\f_B)f^{-1/2}e^{A}\(pS^a_t-i\o \t^a\),\label{eqn:Stz-fourier-new} \\
	&\o\dot S^a_t=4e^{-3A}f^{1/2}\o q_I\frak a^I_a+2ipZ(\f_B)f\dot\t^a,\label{eqn:Stzdot-fourier-new} \\
	&\pa_r\left(\S_{IJ}(\f_B) f^{1/2}e^A\dot{\frak a}^J_a\right)+\S_{IJ}(\f_B) f^{-1/2}e^{-A}\o^2 \frak a^J_a=q_I\dot S^a_t,\label{eqn:Maxz-fourier-new}\\
	&\ddot\t^a+\left(3\dot A+\frac12f^{-1}\dot f+Z^{-1}(\f_B)Z'(\f_B)\dot\f_B\right)\dot\t^a+f^{-1}e^{-2A}\o^2\t^a=-i\o pe^{-2A}f^{-1}S^a_t.\label{eqn:axionz-fourier-new}
	\end{align}
\end{subequations}
Multiplying (\ref{eqn:Stzdot-fourier-new}) with $e^{3A}f^{-1/2}$, taking the radial derivative and substituting (\ref{eqn:Maxz-fourier-new}) and \eqref{eqn:axionz-fourier-new} in the resulting expression gives back (\ref{eqn:Stz-fourier-new}), which is therefore not independent either. In order to simplify the remaining three equations we define the quantities 
\be\label{defs}
\Th^a\equiv S^a_t-\frac{i\o}{p} \t^a,\qquad \Om\equiv\o^2-2p^2fZ(\f_B),
\ee
in terms of which the last three equations in (\ref{eqn:Fluctuations-fourier-new}) become 
\begin{subequations} \label{eqn:Fluctuations-fourier-s}
	\begin{align}
	&\O\dot S^a_t=4e^{-3A}f^{1/2}\o^2 q_I\frak a^I_a-2p^2 Z(\f_B)f\dot\Th^a, \label{dotS}\\
	&-q_I \dot S^a_t+\pa_r\left(\S_{IJ}(\f_B) f^{1/2}e^A\dot{\frak a}^J_a\right)= -\o^2\S_{IJ}(\f_B) f^{-1/2}e^{-A}\frak a^J_a ,\label{eqn:Maxz-fourier-s}\\
	&\pa_r\left(Z(\f_B)e^{3A}f^{1/2}\dot\t^a\right)=-i\o p  Z(\f_B)e^Af^{-1/2}\Th^a.\label{eqn:axionz-fourier-s}
	\end{align}
\end{subequations}
Finally, substituting \eqref{dotS} in \eqref{eqn:Maxz-fourier-s} and \eqref{eqn:axionz-fourier-s} leads to the system of coupled equations
\begin{align}
\label{fluctuation-eqs}\boxed{
\begin{aligned}
&\,\pa_r\(\S_{IJ}(\f_B) f^{1/2}e^A\dot{\frak a}^J_a+2p^2q_IZ(\f_B)f\O^{-1}\Th^a\)\\
&\,\hskip0.5cm+\o^2f^{-1/2}e^{-A}\(\S_{IJ}(\f_B)-4fe^{-2A}\O^{-1}q_Iq_J\)\frak a^J_a-2p^2q_I\o^2\O^{-2}\pa_r(Z(\f_B)f)\Th^a=0,\,\,\,\\
&\rule{0cm}{.8cm}\pa_r\(Z(\f_B)f\O^{-1}\(e^{3A}f^{-1/2}\dot\Th^a-4q_I\frak a^I_a\)\)+Z(\f_B) e^Af^{-1/2}\Th^a=0.
\end{aligned}}
\end{align}
Solving these coupled equations completely determines all fluctuations since $S^a_t$ is related to $\Th^a$ and $\frak a^I_a$ through \eqref{dotS}, and $\t^a$ is subsequently determined from the defining relation for $\Th^a$ in \eqref{defs}. In particular, combining \eqref{defs} and \eqref{dotS} gives 
\be\label{tau}
\dot\t^a=\frac{i\o}{2pZ(\f_B)f}\(4e^{-3A}f^{1/2}q_I\frak a^I_a-\O^{-1}\Big(4e^{-3A}f^{1/2}\o^2 q_I\frak a^I_a-2p^2 Z(\f_B)f\dot\Th^a\Big)\).
\ee

\section{Radial Hamiltonian formalism in the dual frame}
\label{ham}
\setcounter{equation}{0}

In order to construct the holographic dictionary for asymptotically hyperscaling violating Lifshitz solutions in section \eqref{dictionary}, we make use of the radial Hamiltonian formulation of the theory described by the dual frame action \eqref{action}. In this appendix we summarize the essential ingredients of the radial Hamiltonian formalism for the dual frame action \eqref{action}. The exposition is identical to that in section 2 of \cite{Chemissany:2014xsa}, except that (besides fixing a typo) here we include $d_s$ axions and an unspecified number of gauge fields.   

The first step in the Hamiltonian formalism is to decompose the metric and the gauge fields as
\be\label{ADM-metric}
d\bar s^2=(\lbar N^2+\lbar N_i\lbar N^i)d r^2+2\lbar N_id r dx^i+\lbar\g_{ij}(r,x)dx^idx^j,\qquad A^I_\m dx^\m=A^I_r dr+A^I_i dx^i,
\ee
in terms of the lapse and shift functions, respectively $\lbar N$ and $\lbar N_i$, the induced metric $\lbar \g_{ij}$ on the radial slices $\S_r$, as well as the longitudinal and transverse components of the gauge fields. The action \eqref{action} can then be written as an integral
\be
S_\x=\int dr L_\x,
\ee  
over the radial Lagrangian
\bal\label{lagrangian}
L_\x=&\;\frac{1}{2\k^2}\int d^{d_s+1}x\sqrt{-\lbar\g}\;\lbar N\Bigg(\Big(1+\frac{d_s^2\x^2}{\a_\x}\Big) \lbar K^2-\lbar K^{ij}\lbar K_{ij} -\frac{\a_\x}{\lbar N^2}\Big(\dot\f-\lbar N^i\pa_i\f-\frac{d_s\x}{\a_\x}\lbar N \;\lbar K\Big)^2
\NO\\
&-\frac{2}{\lbar N^2}\S_{IJ}^\x(\f)(F^I_{ri}-\lbar N^kF^I_{ki})(F^J_{r}{}^i-\lbar N^lF^J_{l}{}^i)
-\frac{1}{\lbar N^2}Z_\x(\f)\(\dot\c^a-\lbar N^i\pa_i\c^a\)^2\NO\\
&+R[\lbar \g]-\a_\x\pa_i\f\lbar\pa^i\f-\S^\x_{IJ} F^I_{ij}\lbar F^{J\;ij}-Z_\x\pa_i\c^a\lbar\pa^i\c^a-V_\x-2\square_{\lbar\g}\Bigg)e^{d_s\x\f},
\eal
where $\lbar K_{ij}$ is the extrinsic curvature,
\be
\lbar K_{ij}=\frac{1}{2\lbar N}\left(\dot{\lbar\g}_{ij}-\lbar D_i\lbar N_j-\lbar D_j\lbar N_i\right),
\ee
$\lbar K\equiv \lbar\g^{ij}\lbar K_{ij}$ denotes its trace, and $\lbar D_i$ is the covariant derivative with respect to the metric $\lbar\g_{ij}$. 

From the radial Lagrangian \eqref{lagrangian} follow the canonical momenta  
\bal\label{momenta}
\lbar \p^{ij}=&\;\frac{\d L}{\d\dot{\lbar\g}_{ij}}=\frac{1}{2\k^2}\sqrt{-\lbar \g}\;e^{d_s\x\f}\Big(
\lbar K\lbar \g^{ij}-\lbar K^{ij}+\frac{d_s\x}{\lbar N}\lbar\g^{ij}(\dot\f-\lbar N^k\pa_k\f)\Big),\NO\\
\lbar\p^{i}_I=&\;\frac{\d L}{\d\dot A_{i}^I}=-\frac{1}{2\k^2}\sqrt{-\lbar\g}\;e^{d_s\x\f}\S^\x_{IJ}(\f)
\frac{4}{\lbar N}\lbar\g^{ij}(F^J_{rj}-\lbar N^kF^J_{kj}),\NO\\
\lbar \p_\f=&\;\frac{\d L}{\d\dot\f}=\frac{1}{2\k^2}\sqrt{-\lbar\g}\;e^{d_s\x\f}
\Big(2d_s\x \lbar K-\frac{2\a_\x}{\lbar N}(\dot\f-\lbar N^i\pa_i\f)\Big),\NO\\
\lbar\p_{\c a}=&\;\frac{\d L}{\d\c^a}=-\frac{1}{2\k^2}\sqrt{-\lbar\g}\;e^{d_s\x\f}Z_\x(\f)
\frac{2}{\lbar N}(\dot\c^a-\lbar N^i\pa_i\c^a),
\eal
while the momenta conjugate to $\lbar N$, $\lbar N_i$, and $A_r$ vanish identically, and so these fields are non dynamical. Inverting these expressions for the canonical momenta gives Hamilton's equations
\bal\label{inverted-momenta}
\dot{\lbar\g}_{ij}=&\;-\frac{4\k^2}{\sqrt{-\lbar\g}}e^{-d_s\x\f}\lbar N\left(\lbar\p_{ij}-\frac{\a_\x+d^2_s\x^2}{d_s\a}\lbar\p\lbar\g_{ij}-\frac{\x}{2\a}\lbar\p_\f\lbar\g_{ij}\right)+\lbar D_i\lbar N_j+\lbar D_j\lbar N_i,\NO\\
\dot A_i^I=&\;-\frac{\k^2}{2}\frac{1}{\sqrt{-\lbar\g}}e^{-d_s\x\f}\S_\x^{IJ}(\f)\lbar N\lbar\p_{iJ}+\pa_iA_r^I+\lbar N^kF^I_{ki},\NO\\
\dot\f=&\;-\frac{\k^2}{\a}\frac{1}{\sqrt{-\lbar\g}}e^{-d_s\x\f}\lbar N(\lbar\p_\f-2\x\lbar\p)+\lbar N^i\pa_i\f,\NO\\
\dot\c^a=&\;-\frac{\k^2}{\sqrt{-\lbar\g}}e^{-d_s\x\f}Z^{-1}_\x(\f)\lbar N\lbar\p_{\c a}
+\lbar N^i\pa_i\c^a,
\eal
which can be used to determine the radial Hamiltonian of the theory, namely
\be
H=\int d^{d_s+1}x(\dot{\lbar\g}_{ij}\lbar\p^{ij}+\dot A^I_i\lbar\p^i_I+\dot\f\lbar\p_\f+\dot\c^a\lbar\p_{\c a})-L=\int d^{d_s+1}x(\lbar N\ch+\lbar N_i\ch^i+A_r^I\cg_I),
\ee
where $\ch$, $\ch^i$ and $\cg_I$ are given by
\bal
\label{constraints}
\ch=&\;-\frac{\k^2}{\sqrt{-\lbar \g}}e^{-d_s\x\f}\Big(2\lbar\p^{ij}\lbar\p_{ij}-\frac{2}{d_s}\lbar\p^2+\frac{1}{2\a}\Big(\lbar\p_\f-2\x\lbar\p\Big)^2
+\frac14\S^{IJ}_\x\lbar\p^i_I\lbar\p_{iJ}+\frac12Z_\x^{-1}(\lbar\p_{\c a})^2\Big)\NO\\
&\rule{0.8cm}{0cm}+\frac{\sqrt{-\lbar\g}}{2\k^2}\Big(-R[\lbar\g]+\a_\x\lbar\pa^i\f\pa_i\f+\S^\x_{IJ} \lbar F^{Iij}F^J_{ij}+Z_\x\pa_i\c^a\lbar\pa^i\c^a+V_\x+2\square_{\lbar\g}\Big)e^{d_s\x\f},\NO\\
\ch^i=&\;-2\lbar D_j\lbar\p^{ji}+\lbar F^{Ii}{}_j\lbar\p^j_I+\lbar\p_\f\lbar\pa^i\f+\lbar\p_{\c a}\lbar\pa^i\c^a,\NO\\
\cg_I=&\;-\lbar D_i\lbar\p^i_I.
\eal
Since the canonical momenta conjugate to the fields $\lbar N$, $\lbar N_i$, and $A_r$ vanish identically, the corresponding Hamilton equations impose the first class constraints
\be\label{constraints0}
\ch=\ch^i=\cg_I=0,
\ee
reflecting the diffeomorphism and local gauge symmetries of the bulk theory. 

Finally, Hamilton-Jacobi theory provides an alternative expression for the canonical momenta as gradients of Hamilton's principal function $\cs[\lbar\g,A^I,\f,\c]$, namely
\be\label{HJ-momenta}
\lbar\p^{ij}=\frac{\d \cs}{\d\lbar\g_{ij}},\quad \lbar\p^i_I=\frac{\d\cs}{\d A_i^I},\quad \lbar\p_\f=\frac{\d\cs}{\d\f},\quad \lbar\p_{\c a}=\frac{\d\cs}{\d\c^a}.
\ee
Inserting these expressions for the momenta in the constraints \eqref{constraints} one obtains a set of first order functional partial differential equations, known as the Hamilton-Jacobi equations, for the functional $\cs$. Given a solution $\cs$ of the Hamilton-Jacobi equations, Hamilton's equations \eqref{inverted-momenta} become a set of first order equations for the dynamical fields $\lbar\g_{ij}$, $A_i^I$, $\f$ and $\c$. In the radial gauge
\be\label{dual-FG}
\lbar N=e^{-\x\f},\quad \lbar N_i=0,\quad A_r=0,
\ee
which corresponds to Fefferman-Graham gauge in the {\em Einstein} frame, these first order equations take the form
\bal
\label{flow-eqs}
&\dot{\lbar\g}_{ij}=-\frac{4\k^2}{\sqrt{-\lbar\g}}e^{-(d_s+1)\x\f}\left(\left(\lbar\g_{ik}\lbar\g_{jl}-\frac{\a_\x+d^2_s\x^2}{d_s\a}\lbar\g_{ij}\lbar\g_{kl}\right)\frac{\d}{\d\lbar\g_{kl}}-\frac{\x}{2\a}\lbar\g_{ij}\frac{\d}{\d\f}\right)\cs,\NO\\
&\dot A_i^I=-\frac{\k^2}{2}\frac{1}{\sqrt{-\lbar\g}}e^{-(d_s+1)\x\f}\S_\x^{IJ}(\f)
\lbar\g_{ij}\frac{\d}{\d A_j^J}\cs,\NO\\
&\dot\f=-\frac{\k^2}{\a}\frac{1}{\sqrt{-\lbar\g}}e^{-(d_s+1)\x\f}
\left(\frac{\d}{\d\f}-2\x\lbar\g_{ij}\frac{\d}{\d\lbar\g_{ij}}\right)\cs,\NO\\
&\dot\c^a=-\frac{\k^2}{\sqrt{-\lbar\g}}e^{-(d_s+1)\x\f}Z^{-1}_\x(\f)\frac{\d\cs}{\d\c^a}.
\eal

\addcontentsline{toc}{section}{References}


\bibliographystyle{jhepcap}
\bibliography{Lifrefs}

\end{document}